\documentclass[12pt]{iopart}
\usepackage{amssymb}
\usepackage{soul}
\usepackage{color}
\usepackage{epsfig,graphicx}
\usepackage{subfigure}
\graphicspath{{FIGURES_NEW/}}
\usepackage{dcolumn}% Align table columns on decimal point
\usepackage{bm}% bold math
\usepackage{isomath}
\usepackage{epsfig}
\usepackage{ar}

\begin{document}

\title{Role of gravity or confining pressure and contact stiffness in granular rheology}

\author{Abhinendra Singh, Vanessa Magnanimo, Kuniyasu Saitoh and  Stefan Luding }
\address{Multi Scale Mechanics (MSM), MESA + , CTW, University of Twente, Post Office Box 217, 7500 AE Enschede, The Netherlands}

\ead{a.singh-1@utwente.nl}
\begin{abstract}

The steady shear rheology of granular materials is investigated in slow quasi-static states and inertial flows. 
The effect of the gravity field and contact stiffness, which are conventionally trivialized is the focus of this study. 
Series of Discrete Element Method simulations are performed on a weakly frictional granular
 assembly in a split-bottom geometry considering various gravity fields and contact stiffnesses.  While traditionally the inertial number, i.e., the ratio of
 stress to strain-rate timescales describes the flow rheology, we find that a second dimensionless number,
  the ratio of softness and stress timescales, must also be included to characterize the bulk flow behavior.
 For slow, quasi-static flows, the density increases while the macroscopic friction decreases with respective increase in particle softness and gravity.
 This trend is added to the $\mu(I)$ rheology and can be traced back to the anisotropy
 in the contact network, displaying a linear correlation between macroscopic friction and deviatoric  fabric in the steady state.
 Interestingly, the linear relation holds when the external rotation rate is increased for a given gravity field and contact stiffness.

% Various gravity fields and contact
%  stiffnesses are considered. where the response is conventionally assumed to be independent of either.
% % better phrasing needed 
% In previous works, effects of gravity and contact stiffness is believed to play no significant role in influencing the flow behavior of granular materials.
\end{abstract}

%Look for PACS
\section{Introduction}

\label{intro}

Gravity is a critical factor in many natural (granular) phenomena like avalanches, landslides, sand-piles and even in some industrial applications \cite{jaeger1996granular,duran2000sands}.
  Avalanches and debris flows play an important role in the transportation of mass existing at the surface of earth. Gravity-driven flows
 have also been observed on other planetary bodies of our Solar System and are of
 particular importance in understanding the geology of other planets and asteroids as well as for the human exploration of the Moon and Mars in the coming decades \cite{krohn2014mass}. Currently, surface features found on Mars \cite{shinbrot2004dry}, Venus \cite{malin1992mass}, and the Moon \cite{howard1973avalanche} are hypothesized to be the results of avalanches of granular material.

One of the important aspects of granular shear flows is the dependence of stress on external driving.
 Various experimental and numerical studies have shown that for slow--dense, quasi-static flows, the ratio of shear to compressive stress ({\em effective friction coefficient}) is independent of the imposed driving rate \cite{tardos2003slow,campbell2002granular,andreotti2013granular}.
 However, very little is known regarding the same in the presence of very weak gravity fields or low confining stress.
 Shear tests performed on parabolic flights have shown an increase in the friction coefficient
 at low confinement \cite{macari1991analysis,costes1987microgravity,white1990dynamic,alshibli2000constitutive,alshibli1996mechanics,sture1998mechanics}.
 Brucks \etal \cite{brucks2007behavior} also obtained similar trend using centrifuge experiments at gravity levels larger than Earth's gravity.
 Despite these studies, the effect of external compression (gravity) on granular flows is still poorly understood, which leads to issues like
 exploration vehicles getting stuck in the Martian soil.

Soft materials like hydrogel and elastomer, which can support large deformation are of increasing importance
 in engineering and biological applications such as tissue scaffolding,
 biosepration and micro-and-nano-- printing \cite{hoffman2002hydrogels}. 
 While inertial number has been relatively successful in understanding the dynamics of rigid particles \cite{midi2004dense},
 elasticity becomes relevant for soft particles \cite{campbell2002granular,otsuki2010behavior}.
 The deformability of the soft particles has been shown to affect the force network close to the jamming transition \cite{makse2000packing}.
Recent study by Vaart \etal \cite{van2013rheology} has shown different rheological behavior for hard and soft particle suspensions.
Despite the increasing importance, the models for soft deformable particles have been largely ignored.

We claim here that these two factors, i.e., gravity and softness are two aspects of the same phenomenon.
 We aim to test this claim by answering the following questions: 
 (1) How does gravity and softness of particles affect the bulk flow behavior?
 (2) Is there a unique law that can describe the flow behavior on Earth, Mars and the Moon for both soft and rigid particles?

In this paper, we address the above questions with a focus on dense, frictional, quasi-static granular flow.
 Using Discrete Element Method (DEM), we simulate cohesionless frictional granular material in a split-bottom ring shear cell.
%Reference?
 An important aspect of this setup is that the shear rate is given solely by external rotation rate and the geometry.
At the same time, in this geometry the local strain rate does not depend strongly on the external compression \cite{ries2007shear},
 unlike the inclined plane and rotation drum where gravity has a strong effect on the local strain rate  \cite{brucks2007behavior,silbert2007rheology,arndt2006creeping}.
 To study the effect of gravity and particle softness, we independently vary both parameters by two orders of magnitude.
 A change in particle softness provides an adjustment on the microscopic scale, while gravity is a macroscopic (field) modification.
 We find that they have similar effect at the mesoscopic (local) scale.
The bulk behavior can be described well using a dimensionless parameter, defined as the ratio between the time scales due to gravitational compression and contact stiffness.
Further, by increasing the external rotation rate, we study the dependence of macroscopic friction and contact network anisotropy (deviatoric fabric) on the inertial number.
The dependence of macroscopic friction and deviatoric fabric on pressure is added to $\mu(I)$ rheology. Additionally, we find some non-local effects in our results due to
 the presence of gradients in both stress and strain rate, which are quantified by following an approach similar to Koval \etal \cite{KovalRouxCorfdirChevoir2009}. 

The outline of this paper is as follows:
In section \ref{sec:dem} we explain our numerical setup and methodology.
 We present our results for quasi-static state in section \ref{sec:quas}. In section \ref{sec:iner}, we provide results on the rheology and combine it with the results from
 quasi-static state to present new rheological laws.
We then close the paper with a discussion and conclusion of our results in section  \ref{sec:conc} along with a possible outlook for future research.

\section{Discrete Element Method (DEM)}
\label{sec:dem}
We present our numerical simulation scheme and setup in sections \ref{sub:model} and \ref{sub:setup} respectively.
 Section \ref{sec:protocol} briefly presents our averaging methodology and definitions of the tensorial quantities.
We summarize various time scales associated with the system in section \ref{sub:param-diffg}.

\subsection{Model}
\label{sub:model}
Our computational techniques are based on the soft-sphere DEM simulations as developed by Cundall and Strack \cite{cundall1983modeling}, Walton \cite{walton1993numerical} and Luding
 \cite{luding2001from_1,luding2008introduction,luding2008cohesive}.
The normal force between the particles in contact is modeled by a Hookean spring as $f_n=-k_n\delta_n-\eta_n v_n$,
where $k_n$, $\delta_n$, $\eta_n$, and $v_n$ are the normal stiffness, particle overlap, normal viscous damping coefficient, and relative velocity in the normal direction, respectively.
Similarly, the tangential force is modeled as $f_t=-k_t\delta_t-\eta_t v_n$,
where $k_t=2k_n/7$, $\delta_t$, $\eta_t=\eta_n/4$, and $v_t$ are the tangential stiffness, relative displacement in tangential direction,
tangential viscous damping coefficient, and relative velocity in tangential direction, respectively.
We also introduce Coulomb's friction between the particles, where the tangential force $f_t$ is switched to the sliding force
$f_s=-\mu_p|f_n|$, $\mu_p$ being the particle friction coefficient when $f_t$ exceeds the critical value,\ i.e.\ $|f_t|>\mu_p|f_n|$ (with $\mu_p=0.01$) \cite{luding2008cohesive}.

To study the effect of particle softness on macroscopic behaviors, we explore a range of normal contact stiffness $k_n$, from $10\ \mathrm{Nm^{-1}}\leq k_n\leq10^4\ \mathrm{Nm^{-1}}$.
 While changing $k_n$, the time step for numerical integration $\delta t$
 is adjusted to be $\frac{1}{50}$ times the contact duration to ensure accurate dynamic integration \cite{luding2008cohesive}. When $k_n$ is changed, $k_t$ and $\eta_n$ are changed
 as well, to ensure that the coefficient of restitution remains unchanged.

\subsection{Split-bottom ring shear cell}
\label{sub:setup}
%
%%%%%%%%%%%%%%%%%%%%%%%%%%%%%%%%%%%%%%%%%%%%%%%%%%%%%%%%%%%%%%%%%%%%%%%%%%
\begin{figure}
\centering
\includegraphics[width=6cm]{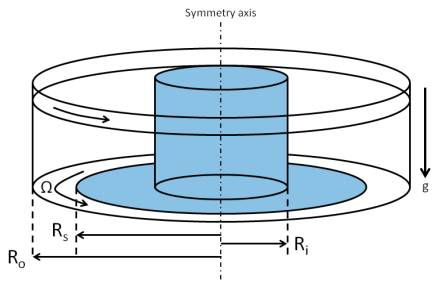}
\caption{(Color online)
A sketch of our numerical setup consisting of a fixed inner part (light blue shade) and a rotating outer part (white).
The white part of the base and the outer cylinder rotate with the same angular velocity, $\Omega$, around the symmetry axis (dot-dashed line).
The inner, split, and outer radii are given by $R_i=8\langle a\rangle$, $R_s=40\langle a\rangle$, and $R_o=52\langle a\rangle$, respectively,
where each radius is measured from the symmetry axis. The gravity, $g$, points downwards as shown by an arrow.}
\label{fig:sketch_sp}
\end{figure}
%%%%%%%%%%%%%%%%%%%%%%%%%%%%%%%%%%%%%%%%%%%%%%%%%%%%%%%%%%%%%%%%%%%%%%%%%%
%
Figure \ref{fig:sketch_sp} is a sketch of our numerical setup \cite{fenistein2003kinematics,fenistein2004universal,luding2008effect,dijksman2010granular}. In this figure,
 the inner, split, and outer radii are given by $R_i$, $R_s$, and $R_o$, respectively, where the concentric cylinders rotate relative to each other at a rate $\Omega$ around the symmetry axis (the dot-dashed line).
 The ring shaped split at the bottom separates the
 moving and static parts of the system, where a part
 of the bottom and the outer cylinder rotate at the
 same rate. The system is filled with $N\approx 3.7\times10^{4}$ polydispersed spherical particles with
density $\rho= 2000\ \mathrm{kgm^{-3}}=2\ \mathrm{gcm^{-3}}$
 up to height $H$.
The average size of particles is $\langle a \rangle=1.1$\,mm, and the width of the homogeneous 
size-distribution (with $a_{\rm min}/a_{\rm max}=1/2$) is $1-{\cal A} = 1-\langle a \rangle^2/\langle a^2 \rangle = 0.18922$.
The cylindrical walls and the bottom are roughened due to some (about $3\%$ of the total number) attached/glued particles \cite{luding2008effect,luding2008,Singh2014effect}.
The initial state of the system is prepared frictionless. Friction is switched on at the onset of shear.

When there is relative motion at the split, a shear band initiates at the bottom from the split position $R_s$
 and propagates upwards and inwards, remaining far away from the cylindrical walls in most cases \cite{fenistein2004universal,dijksman2010granular}.
 The qualitative behavior is governed by the ratio $H/R_s$ and three regimes can be observed as reported in \cite{fenistein2004universal,dijksman2010granular}.
 We keep $H/R_s< 0.5$, such that the shear band reaches the free surface and stays away from inner wall.

To understand the effect of gravity on the bulk behavior, gravity is varied
 in the range $0.5\ \mathrm{ms^{-2}}\le g\le50\ \mathrm{ms^{-2}}$. The details regarding rotation
 rate of the system are reported in table \ref{tab:time_scales}.
 The total simulation time is chosen such that the
 cylinder completes half a rotation in that time.

\subsection{Local averaging}\label{sec:protocol}
One of the goals of current research in the granular community is to derive  
macroscopic continuum theory based on the given micro-mechanical properties \cite{bagi1999microstructural,latzel2000macroscopic,weinhart2013coarse}. 
We assume translational invariance in the azimuthal $\theta-$direction. The averaging
 is thus performed over toroidal volumes over many snapshots in time, leading to
 fields $Q(r,z)$ as function of the radial and vertical positions. Here, the averaging is performed
 with spacings $\Delta r$ and $\Delta z$ around two particles diameter in radial and vertical directions
(averaging procedure for two and three dimensions
is discussed in detail in Refs.\ \cite{luding2008,latzel2000macroscopic} and hence not discussed further here).

\subsubsection{Macroscopic (tensorial) quantities}\label{sec:tensor}
Here, we present general definitions of the averaged
microscopic and macroscopic quantities
 -- including strain rate, stress and fabric (structure)
 tensors.

The strain rate is calculated by averaging velocity gradients of particles over cells and is given by
\begin{equation}
 \dot{\epsilon}_{\alpha\beta}(\mathbf{r})=\frac{1}{2}\displaystyle\sum_{p \in V}\left( \bigtriangledown_\beta {\rm{v}}^p_\alpha+\bigtriangledown_\alpha {\rm{v}}^p_{\beta}\right) ~,
\end{equation}
 where Greek letters represent components radial distance $r$, azimuthal angle $\theta$, and height $z$, while $\bigtriangledown$ represents the gradient
 in cylindrical coordinate \cite{luding2008}.

The stress tensor is given by 
\begin{equation}
  {{\sigma}_{\alpha\beta}}(\mathbf{r})=\displaystyle\frac{1}{V} \left[\displaystyle\sum_{p \in V}{{  m^p}({  \mathrm {v}^p_\alpha})({  \mathrm {v}^p_\beta})}-
\displaystyle\sum_{c \in V}{{  l^c_{\alpha}}{  f^c_{\beta}}} \right] ~,
 \label{eq:Stress}
\end{equation}
 with particles $p$, contacts $\mathrm c$, mass $\mathrm  m^p$, velocity $\mathrm v^p$, force $  f^c$ and branch vector $  l^c$,
 while Greek letters represent components $r$, $\theta$, and $z$ \cite{luding2008}.
The first term is the sum of kinetic energy fluctuations, and the second term involves the 
dyadic product of the contact-force with the contact-branch vector.

The quantity which describes the local configuration of a granular assembly is the fabric tensor 
 \cite{imole2012hydrostatic,weinhart2013coarse} and is given by
 \begin{equation}
  {F}_{\alpha\beta}(\mathbf{r})=\displaystyle\frac{1}{V}\displaystyle\sum_{p\in V}V^{p}\displaystyle\sum_{c\in p}{  n^{c}_\alpha}{  n^{c}_\beta} ~,
 \label{eq:Fabric}
 \end{equation}
where $V^p$ represents the particle volume which lies inside the averaging volume $V$, $n^c$ is the normal unit branch-vector 
pointing from the center of particle $p$ to contact $c$.

\subsubsection{Isotropic and deviatoric parts}\label{sec:tensor-iso-dev}
Any given tensor $\boldsymbol{Q}$ can be uniquely decomposed into isotropic and deviatoric parts as

\begin{equation}
 \boldsymbol{Q} =Q_V\boldsymbol{I}+\boldsymbol{Q_D}
\label{eq:ten-split}
\end{equation}
with $Q_V=\frac{1}{3} \Tr {\boldsymbol{Q}}$ and the traceless deviator $\boldsymbol{Q_D}$. The latter contains information about the eigensystem of $\boldsymbol{Q}$,
 that is identical to the eigensystem of $\boldsymbol{Q_D}$ itself.

Let us assume $Q_1$, $Q_2$, and $Q_3$ are the eigenvalues of $\boldsymbol{Q_D}$ sorted such that  $Q_1\ge Q_2\ge Q_3$.
Based on the normalization, we use following definition to quantify the anisotropy of any tensor $\boldsymbol{Q_D}$ using a single scalar quantity:

\begin{equation}\label{eq:qdev}
 Q_{\rm dev}=\sqrt{\frac{{((Q_{1}-Q_{2})^2+(Q_{2}-Q_{3})^2+(Q_{3}-Q_{1})^2)}}{6}} ~,
\end{equation}
 the deviators $\dot{\epsilon}_{\mathrm{dev}}$, $\sigma_{\mathrm{dev}}$, and $F_{\mathrm{dev}}$
 refer to strain rate $\dot{\epsilon}_{\alpha\beta}$, stress $\sigma_{\alpha\beta}$ and fabric $F_{\alpha\beta}$ respectively. Other definitions of $Q_{\rm dev}$ are reported in \cite{kumar2014macroscopic}.

$\dot{\gamma}=\dot{\epsilon}_{\mathrm{dev}}$ quantifies the local strain rate magnitude,
 which is very close to the form defined in cylindrical coordinates \cite{depken2006continuum} as tested in \cite{luding2008}.
The pressure $p$ is the isotropic hydrostatic stress, while $\tau=\sigma_{\rm dev}$ quantifies objectively the
shear stress. The deviatoric stress ratio, $\mu={\tau}/{p}$, quantifies the \textquotedblleft stress anisotropy\textquotedblright or the macroscopic friction coefficient.
The volumetric fabric $3 F_{\rm v}$ represents the contact number density, while 
the deviatoric fabric $F_{\rm dev}$ quantifies the anisotropy of the contact network (as studied in detail in \cite{imole2014micro}).

%% Tables

\fulltable{\label{tab:time_scales}Table showing the values of $g$ (units of $\rm ms^{-2}$) and particle stiffness $k_n$ (units of $\rm Nm^{-1}$) used in our simulations,
and various time scales associated with the system, as discussed in the main text (in units of $\rm s$).
The values of $T_{\dot{\gamma}}$ and $T_p$  are the average values reported at $z=2\langle d\rangle,H/2,H-2\langle d\rangle$
 in the center of the shear band.}
\br
$g$ & $\frac{\Omega}{2\pi}$ & $k_n$  & $T_c\times10^{-3}$  & $T_{\eta}\times10^{-3}$ & $T_g\times10^{-2}$ & $T_{\dot{\gamma}}$ &  $T_p\times10^{-3}$ \\
\mr
% Lines of table here ending with \\
$0.5$ &0.005  &100      & $2$ & $5.6$   & $6.6$ & 25, 20, 10 & ($1.7$, $2.5$, $5$)    \\
\mr
$1$ &0.01  &100      & $2$ & $5.6$   & $4.7$ & 10.9, 7.8, 2.7 & ($32$, $15.3$, $12.5$) \\
\mr
$2$ &0.01  &100      & $2$ & $5.6$   & $3.3$ & 10.7, 7.5, 2.7 & ($9$, $11$, $22$)   \\
\mr
$5$  &0.01 &100 & $2$ & $5.6$    & $2.1$ & 10.3, 7.4,2.6 & ($5.9$, $7$, $14.6$)    \\
5 &0.01 &500 & $0.1$    & $2.5$ & $2.1$ & 10.6, 7.5, 2.1 & ($5.1$, $7$, $14.1$)   \\
\mr
$20$ & 0.01 &100 & $2$ & $5.6$    & $1.05$ & 9.7, 7.0, 2.6 & ($2.9$, $3.8$, $8$)    \\
20 & 0.01 &400    & $10$ & $11.2$ & $1.05$ & 10, 7.1, 2.7 & ($2.9$, $3.6$, $7.4$)   \\
\mr
$50$ &0.01  &100 & $2$ & $5.6$    & $0.66$ & 8.7, 6.6, 2.5 & ($1.8$, $2.2$, $4$)   \\
50 & 0.01 &1000 & $0.66$  & $18$   & $0.66$ & 10.1, 7.1, 2.6 & ($1.9$, $2.5$, $7$)   \\
\mr
$10$ &0.01  &100 & $2$ & $5.6$   & $1.5$ & 9.9, 7.0, 2.6 & ($4$, $5.6$, $24$)    \\
10 &0.01 &1000 & $0.66$ & $18$     & $1.5$ & 9.1, 8.1, 2.6 & ($4$, $5.6$, $27$)    \\
10 &0.01 &10000 & $0.2$  & $0.56$  & $1.5$ & 10.7, 7.3, 2.8 & ($4$, $5.4$, $31$)   \\
\mr
10 &0.1  &100 & $2$ & $5.6$   & $1.5$ & 1.1, 0.7, 0.23 & ($4$, $6$, $9$)    \\
10 &0.5  &100 & $2$ & $5.6$   & $1.5$ & 0.2, 0.15, 0.05 & ($4$, $5$, $10$)    \\
10 &1.0  &100 & $2$ & $5.6$   & $1.5$ & 0.1, 0.07, 0.02 & ($4$, $5$, $20$)   \\
10 &2.0  &100 & $2$ & $5.6$   & $1.5$ & 0.02 0.03, 0.008 & ($4$, $6$, $18$)    \\
\br
\endfulltable

\fulltable{\label{tab:dim_num}
Table showing the values of $g$ (units of $\rm ms^{-2}$) and particle stiffness $k_n$ (units of $\rm Nm^{-1}$) used in our simulations,
and various dimensionless numbers as introduced in the text, as discussed in the main text (in units of $\rm s$).
The average values of $I$ are reported at $z=2\langle d\rangle,H/2,H-2\langle d\rangle$
 in the center of the shear band.}
\br
$g$ & $\frac{\Omega}{2\pi}$ & $k_n$    & $I$ & $\kappa^2_g$ \\
\mr
% Lines of table here ending with \\
$0.5$ &0.005  &100    & ($7$, $12$, $50$) $\times10^{-3}$& $2\times10^{-5}$   \\
\mr
$1$ &0.01  &100       & ($1.3$, $2$, $4$) $\times10^{-3}$ &  $5\times10^{-4}$\\
\mr
$2$ &0.01  &100      & ($0.75$, $2$, $1.4$) $\times10^{-3}$ & $3.4\times10^{-3}$ \\
\mr
$5$  &0.01 &100   & ($2$, $0.7$, $0.9$) $\times10^{-3}$ & $1\times10^{-3}$   \\
5 &0.01 &500    &  ($2.5$, $0.8$, $1.4$) $\times10^{-3}$ & $2\times10^{-4}$ \\
\mr
$20$ & 0.01 &100   & ($1$, $0.5$, $0.9$) $\times10^{-3}$ & $1\times10^{-3}$  \\
20 & 0.01 &400     & ($2$, $4$, $7$) $\times10^{-4}$ & $1\times10^{-4}$ \\
\mr
$50$ &0.01  &100 &  ($8$, $3.4$, $7.2$) $\times10^{-5}$ & $2.5\times10^{-3}$ \\
50 & 0.01 &1000 &  ($5$, $3$, $6$) $\times10^{-5}$ & $2.5\times10^{-4}$ \\
\mr
$10$ &0.01  &100 & ($1.1$, $0.6$, $1.5$) $\times10^{-3}$ & $5\times10^{-4}$  \\
10 &0.01 &1000  & ($1.4$, $0.5$, $1.6$) $\times10^{-3}$ & $5\times10^{-5}$  \\
10 &0.01 &10000 & ($1.2$, $0.6$, $0.9$) $\times10^{-3}$ & $5\times10^{-6}$  \\
\mr
10 &0.1  &100 & ($1.5$, $0.6$, $0.9$) $\times10^{-2}$ & $5\times10^{-4}$  \\
10 &0.5  &100   & ($7.5$, $2.6$, $6$) $\times10^{-2}$ & $5\times10^{-4}$  \\
10 &1.0  &100   & ($1.5$, $0.5$, $10.2$) $\times10^{-1}$ & $5\times10^{-4}$  \\
10 &2.0  &100   & 0.3, 0.16, 1.51 & $5\times10^{-4}$  \\
\br
\endfulltable

\subsection{Time scales}
\label{sub:param-diffg}
We characterize the dynamics of the system looking at different time scales.
At first, we define two microscopic time scales related to the contact duration and the viscous damping coefficient between two particles in contact, respectively, as
\begin{equation}
T_k=\sqrt{\frac{\langle m\rangle}{k_n}}~,\hspace{5mm}
T_\eta=\frac{\langle m\rangle}{\eta_n}~,
\label{eq:micro}
\end{equation}
 where $\langle m\rangle$ is the mass of a particle
 with mean diameter. $T_k$ and $T_\eta$ can be related to contact time $T_c=\frac{\pi}{\sqrt{\frac{1}{T^2_k}-\frac{1}{T^2_\eta}}}$.
Next, two time scales associated with external forces,\ i.e.\ the gravity and external rotation rate, can be introduced as
\begin{equation}
T_g=\sqrt{\frac{\langle d\rangle}{g}}~,\hspace{5mm}
T_\Omega=\frac{2\pi}{\Omega}~,
\label{eq:external}
\end{equation}
respectively, where $T_g$ is the time taken by a particle with zero initial velocity to fall a distance $\langle d\rangle/2$.

The time scales, \eref{eq:micro} and \eref{eq:external}, are functions of material constants
and applied external forces, hence, are constants throughout the system.
In this sense, the time scales, $T_c$, $T_\eta$, $T_g$, and $T_\Omega$, are \emph{global}.
On the other hand, two local time scales related to the local  shear rate $\dot{\gamma}$ and pressure $p$, as introduced
 in\ \cite{midi2004dense, da2005rheophysics} are:
\begin{equation}
T_{\dot{\gamma}}=\frac{1}{\dot{\gamma}}~,\hspace{5mm}
T_p=\langle d\rangle\sqrt{\frac{\rho}{p}}~.
\label{eq:macro}
\end{equation}
As shown in the following sections, the spatial distributions of pressure and shear rate are inhomogeneous due to gravity and shear band localization.
Therefore, in contrast to the global time scales, $T_{\dot{\gamma}}$ and $T_p$ are \emph{local} field variables that depend on spatial position.

The time scales can be combined to formulate dimensionless numbers that give indications of dominance of one of the time scales.
For example, the inertial number, as introduced by \cite{da2005rheophysics,JopForterrePouliquen2006,forterre2008flows},
\begin{equation}
  I\equiv T_p/T_{\dot{\gamma}}=\dot{\gamma}\langle d\rangle/ \sqrt{p/\rho} ~,
\end{equation}
provides an estimate of the local rapidity of the flow.
For $I\ll 1$, the flow is \emph{quasi-static}, where particles interact via enduring contacts and inertial effects are negligible.
 For $I\sim 1$, the flow is in the \emph{dense inertial regime}, and for $I\gg 1$, the flow is in the rapid, collisional gas like state.
 A variant of inertial number $I_k=\dot{\gamma}\langle d\rangle/ \sqrt{p_{kin}/\rho}$ can be defined by using only the
 kinetic pressure instead of the total pressure. Other expressions such as Savage or Coulomb number
 ${\dot{\gamma}}^2 {\langle d\rangle}^2\rho/{p}$ is simply the square of the inertial number $I$ \cite{savage1984mechanics}.

A dimensionless parameter \emph{global compressibility} can be introduced as the ratio between $T_g$ and $T_k$:
\begin{equation} 
\kappa_g \equiv \frac{T_k}{T_g} = \sqrt{\frac{\langle m\rangle g}{k_n\langle d\rangle}}~,
\label{eq:kappa}
\end{equation}
 providing a global measure of compressibility of the bulk material.
A high $\kappa_g$ signifies that the bulk material is compressible, which comes from either very high confinement by strong gravity or from low contact stiffness at particle level.
On the other hand, when $\kappa_g$ is small,
 the average overlap is very small compared to the particle diameter, which means that the bulk material is closer to being the rigid limit.
A similar dimensionless parameter can be introduced as:
\begin{equation}
\kappa_p \equiv \frac{T_k}{T_p} = \sqrt{\frac{p \langle d\rangle }{k_n}} ~,
\label{eq:kappa_p}
\end{equation}
 which estimates the compressibility of the material at the local scale.

\Tref{tab:time_scales} shows typical values of timescales in our simulations, and  \tref{tab:dim_num} reports various dimensionless numbers.
 We observe that for flows with a rotation rate $\Omega/2\pi=0.01\ \rm s^{-1}$ and the gravity $g\ge 1\ \rm {ms^{-2}}$, the inertial number $I$ is well below $1$
 for all values of $k_n$.
 For lower values of gravity $g=0.5 \rm ms^{-2}$, the rotation rate is chosen to be $\Omega/2\pi=0.005\ s^{-1}$, such that $I$ stays in the same range.
From this table we infer that the systems for wide a range of $g$ and $k_n$ can be safely assumed to be in the rate-independent quasi-static state.
However, we observe that with increase in rate of rotation $\Omega/2\pi$, $I$ begins to increase and the flow behavior
 enters into the rate dependent inertial regime.

%%%%%%%%%%%%%%%%%%%%%%%%%%%%%%%%%%%%%%%%%%%%%%%%%%%%%%%%%%%%%%%%%%%%%%%%%%
\section{Quasistatic rheology}
\label{sec:quas}
In this section, we present our results on the analysis of macroscopic rheology in the quasi-static state.
At first, in section \ref{sub:stress-sb} we study the steady state and critical state and the amount of time system requires to reach the critical state.
 We explore the effect of gravity and stiffness on the macroscopic friction coefficient in section  \ref{sub:fric}.
We also show the results of local volume fraction in section \ref{sub:volu},
and connect the rheology to the microscopic structure tensor in section \ref{sub:micor}. We will extend our analysis to dense inertial flows in section \ref{sec:iner}.

\subsection{Steady state and critical state}
\label{sub:stress-sb}
 Shearing leads to dilation and a build-up of shear stress and anisotropy in fabric. Additionally, anisotropy
 in the fabric of the granular medium needs a finite amount of strain to build up. Since, we are interested in the steady state
 flow properties, we need to ask: what is the minimum time or equivalently strain required to reach the steady state
 flow regime?

We define $\varphi=\frac{180}{2\pi} \Omega \Delta t$
 (where $\Delta t$ is the simulation time) as a shear parameter, which is the same as the amount, in degrees,
through which the system rotates.
In general, the time required for stabilization can depend on the considered variable \cite{KovalRouxCorfdirChevoir2009}. In the following,
 we consider the relaxation of various quantities to the steady state (both locally and globally averaged). As we expect this relaxation
 to be slowest for small $\Omega$, we perform these tests for $\Omega/2\pi=0.01\ \rm s^{-1}$ that is the smallest $\Omega$ we explore in our simulations.

 At first, we analyze the global quantities (averaged over the whole system) like the averaged kinetic energy and average number of contacts.
 We find that they relax very fast ($\varphi \sim 5$) and are not reliable since they only represent the
 fastest relaxation in the system.

Next, we analyze the relaxation of local quantities. In order to estimate the relaxation time for local quantities, we analyze local velocity profiles.
 We find that such a relaxation requires ($\varphi \sim 30$ or more) which is in agreement with Ries \etal \cite{ries2007shear}.
 Consequently we consider that the condition to approach the steady state
 is $\varphi \ge 30$. Experimental results from Wortel \etal \cite{wortel2014rheology} also show that the system needs to be rotated
 by an angle of approximately 30 degrees in order to reach a reasonable steady state. 
When the steady state is reached, the local averaging is performed over almost 600 snapshots distributed over $\varphi \ge 30$.

The consistency of the local averaged quantities also depends on the accumulated local shear strain during the procurement of data.
 We concentrate our interest in the region where the system can be considered in a critical state, which occurs at large enough shear strain
 $\gamma =\dot{\gamma}\Delta t \ge 1$. In other words, the system can be assumed to be in the critical state.
  The critical state is a unique state reached after long shear, where material deforms with applied strain without any change in normal stress,
 shear stress and volume fraction, and the system forgets its preparation history \cite{wood1990soil}.

\begin{figure}
\centering
\includegraphics[width=5cm,angle=-90]{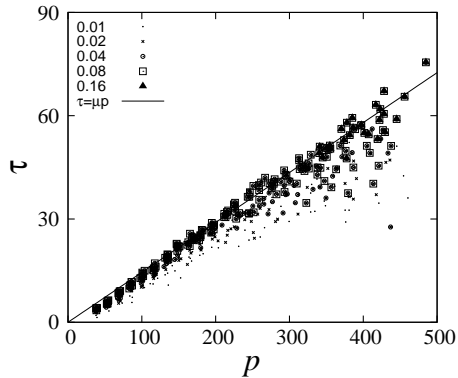}
\caption{(Color online)
The local shear stress, $\tau(r,h)$, plotted against the local pressure, $p(r,h)$,
for different values of the local shear rate, $\dot{\gamma}(r,h)$ as given in the legend (in units of $\mathrm{{s}^{-1}}$), with gravity, $g=10\ \rm{ms^{-2}}$
 and $\Omega/2\pi=0.01\ \rm s^{-1}$.}
\label{fig:tau-p}
\end{figure}

 Figure \ref{fig:tau-p} shows the local shear stress $\tau(r,h)$ plotted against the local pressure $p(r,h)$.
 We observe that for a given pressure, $\tau$ is higher for larger $\dot{\gamma}$,
 however for $\dot{\gamma}>\dot{\gamma}_c$ (with $\dot{\gamma}_c \approx 0.08\ {\rm s^{-1}}$), $\tau$ becomes almost independent of the local strain rate.
 This means that $\tau/p$ is almost constant for all data points with strain rate $\dot{\gamma}>\dot{\gamma}_c$.
 In other words, if the dimensionless shear length $l_{\gamma}=t_{av}\dot{\gamma}$ \cite{luding2008} exceeds unity (where $t_{av}$ is the time over which averaging of
 the data is performed),
 layers can be assumed to be sheared by almost a particle diameter.
 For $l_{\gamma}\ge 1$, the shear deformation can be considered to be fully established,
 and the system is assumed to be in the critical state \cite{luding2008}. 

 In the same setup, Ries \etal \cite{ries2007shear} and  Szab\'{o} \etal \cite{szabo2014evolution} also found that after long enough shear, the material
 inside the shear band reaches the critical state and can be characterized by estimating local accumulated strain $\gamma\ge 1$.
 Our previous works \cite{luding2008,singh2013effect} also showed that for rotation rate $\Omega/2\pi=0.01\ \rm s^{-1}$,
 $\dot{\gamma}_c \approx 0.1\ {\rm s^{-1}}$ is the shear rate above which the shear band is well established.
 Since we are interested in the flow behavior of the material, as default
 in the rest of the paper we focus only on the data well inside the shear band with local strain rate,
\begin{equation}
\dot{\gamma}(r,h)>\dot{\gamma}_c(\Omega)\equiv \frac{10\Omega}{2\pi} ~,
\label{eq:cutoff}
\end{equation}
 unless specified otherwise.

\subsection{Friction coefficient}
\label{sub:fric}
We will now turn our attention towards the effect of gravity and softness on the macroscopic friction coefficient
 $\tau/p$ in the quasi-static state. In previous studies, the friction coefficient has been assumed to be independent of both the particle stiffness and gravity.
 However, particles used in these were extremely rigid.
%% Cite 
 Few studies were performed systematically investigating
 the dependence  of the flow behavior on gravity \cite{macari1991analysis,costes1987microgravity,white1990dynamic,alshibli2000constitutive,alshibli1996mechanics,sture1998mechanics}.

\begin{figure}
\centering
\subfigure[]{\includegraphics[width=5cm,angle=-90]{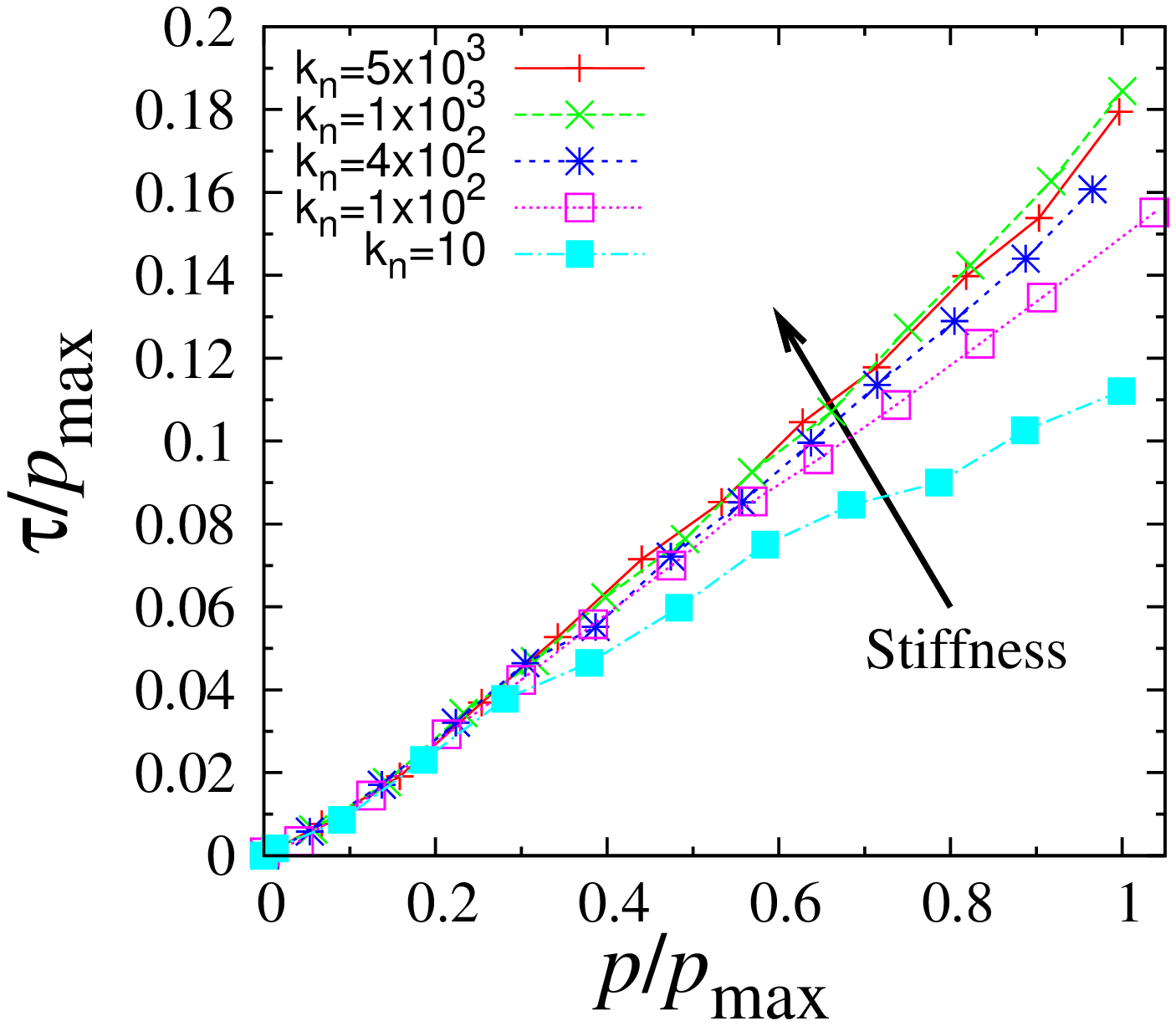}\label{fig:tau-p-k}}
\subfigure[]{\includegraphics[width=5cm,angle=-90]{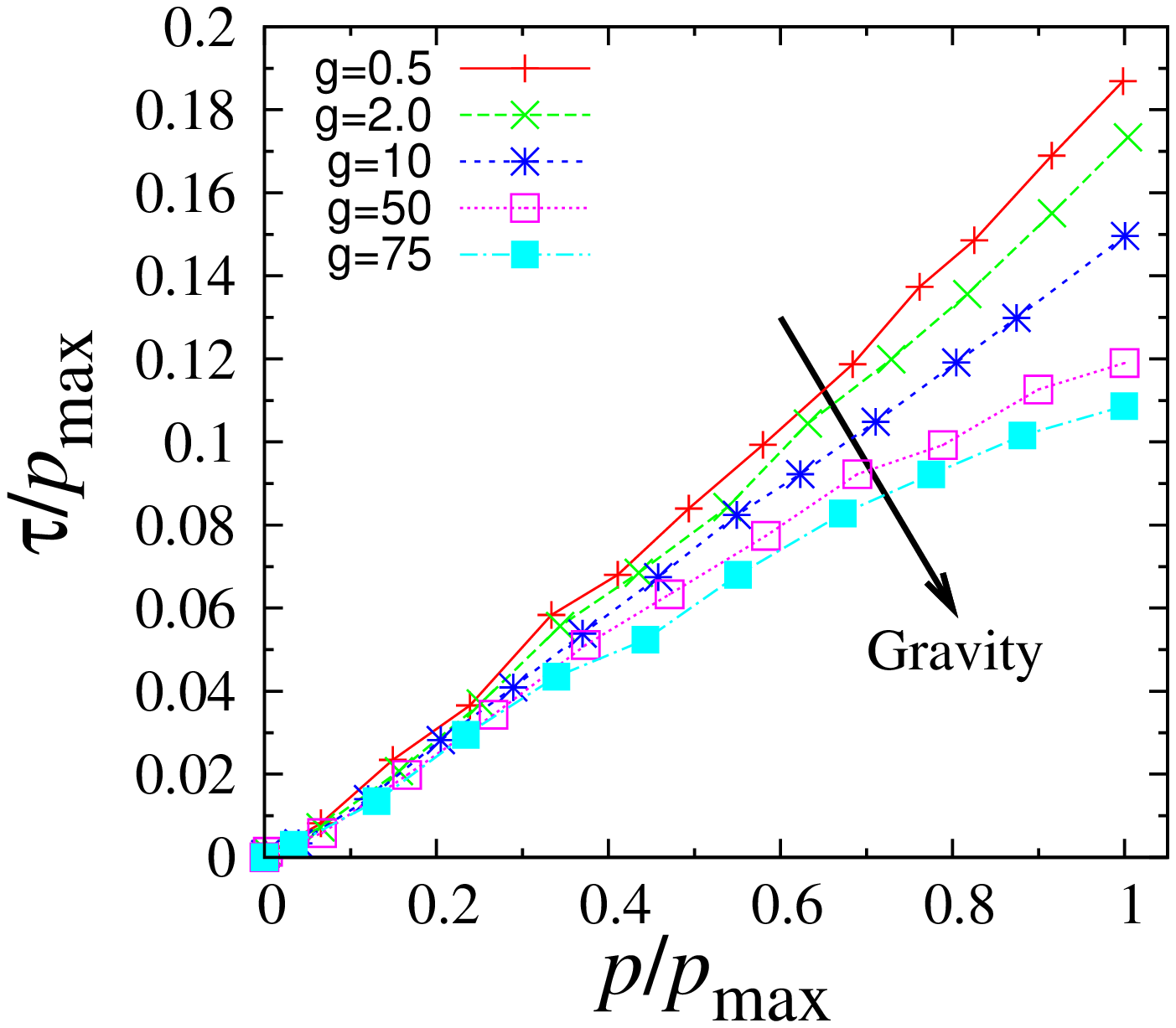}\label{fig:tau-p-g}}
\caption{(Color online)
(Left) The local shear stress plotted against the local pressure for different values of the normal stiffness
as given in the legend in units of $\mathrm{Nm^{-1}}$. Here, the gravity is fixed to $g=10\ \mathrm{ms^{-2}}$.
(Right) The local shear stress plotted against the local pressure for different values of the gravity
as given in the legend in units of $\mathrm{ms^{-2}}$. Here, the normal stiffness is fixed to $k_n=100\ \mathrm{Nm^{-1}}$.
Both $\tau(r,h)$ and $p(r,h)$ are scaled by the maximum pressure $p_\mathrm{max}(r,h)$, respectively.}
\label{fig:tau-p-k-g}
\end{figure}

Figures \ref{fig:tau-p-k} and \ref{fig:tau-p-g} show shear stress-pressure curves for different values
 of normal stiffness $k_n$ and external gravity $g$, respectively. In these plots, for the sake of clarity, both shear stress and pressure
 are plotted only at the center position $R_c$ of the shear band ($R_c$ being the the position at which the local shear rate is maximum).
 For a better comparison, both shear stress and pressure are normalized
 with the maximum pressure $p_{\rm max}$ reached in the simulation with particular $k_n$ and $g$ (so that both abscissa and ordinates are
 of the same order). We observe that both softness of the particles (interpreted as opposite of contact stiffness) and external gravity drastically affect the shear stress. 
Moreover, they act in the same direction as for a given pressure
 $\tau$ decreases with increase in either particle softness or the external gravity. We also see that with increasing softness or gravity,
 the relation between $\tau$ and $p$ becomes non-linear.

\subsubsection{Linear approximation}
\label{subsub:linear}
To understand the dependence of the macroscopic friction coefficient in a quasi-static state on the softness and gravity,
we estimate it as the slope of a linear fitting function for the shear stress against pressure
 in the same fashion as the Mohr-Coulomb failure criterion,\ i.e.\
\begin{equation}
\tau(r,h) \simeq \mu_0^\mathrm{global} p(r,h)~,
\label{eq:barmu}
\end{equation}
where $\mu_0^\mathrm{global}$ is the \emph{global friction coefficient} which depends neither on the shear rate nor on the pressure.

\begin{figure}
\centering
\subfigure[]{\includegraphics[width=5cm,angle=-90]{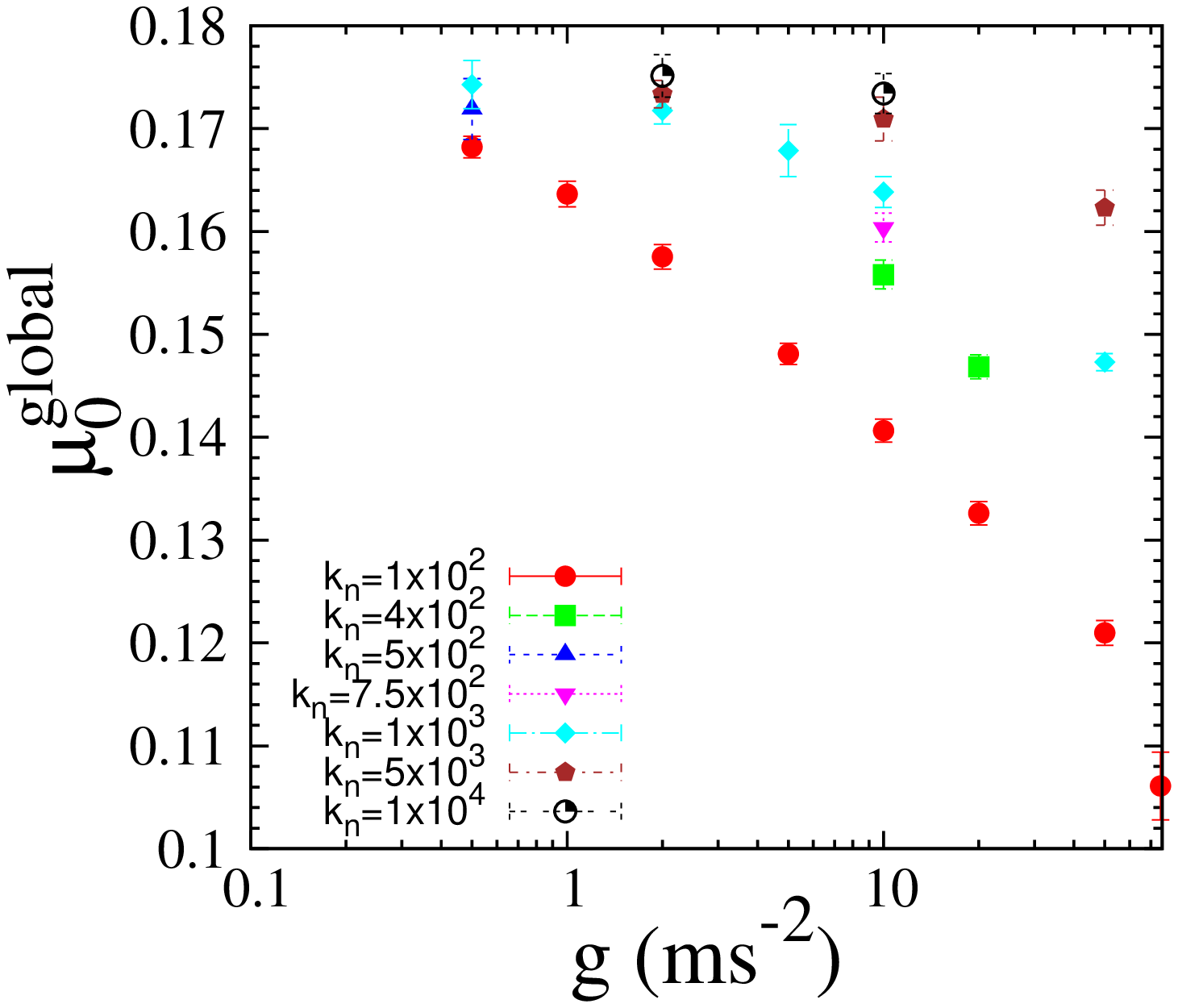}\label{fig:mu-g}}
\subfigure[]{\includegraphics[width=5cm,angle=-90]{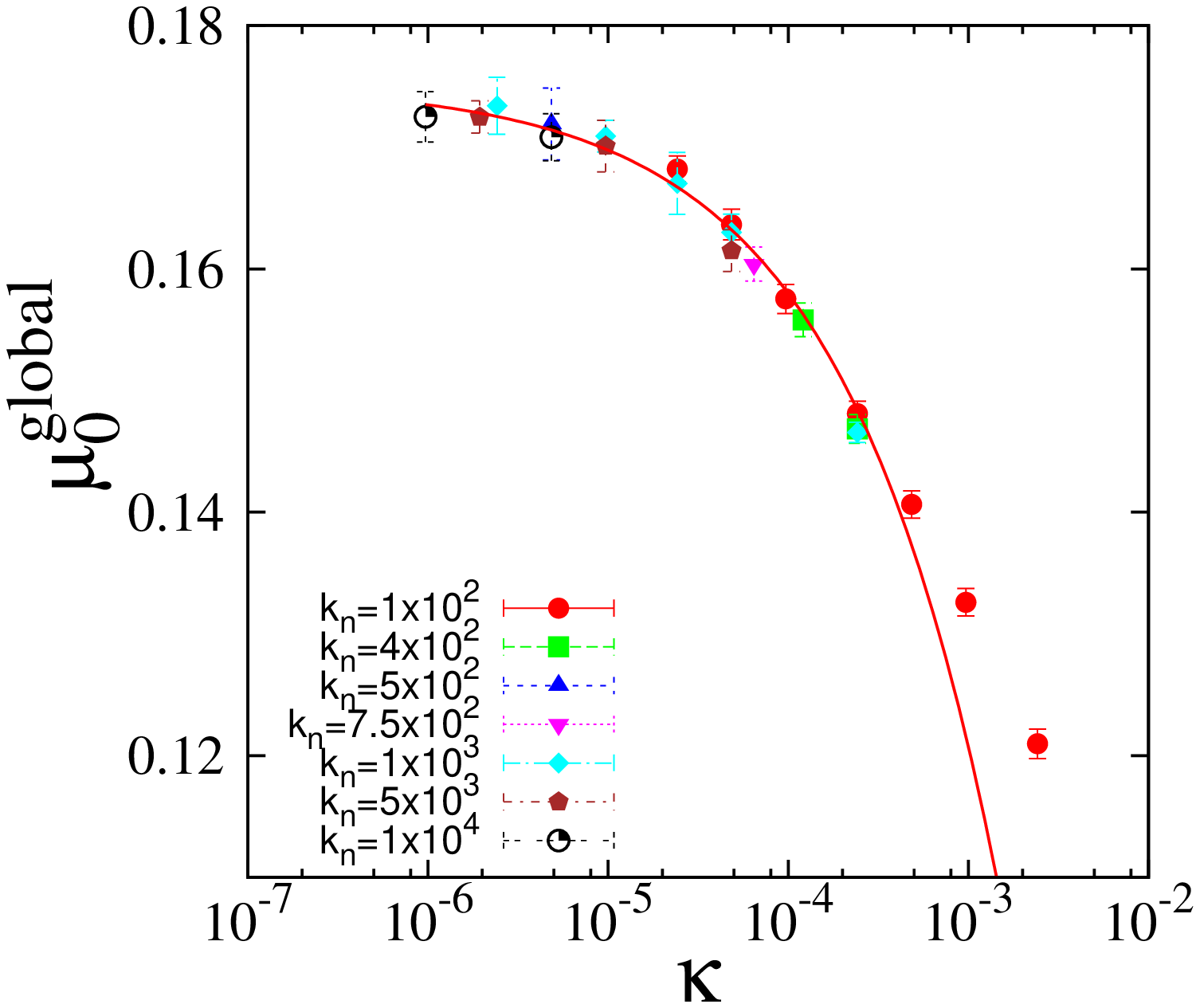}\label{fig:mu_g_kappa}}
\caption{(Color online)
The global friction coefficient, $\mu_0^\mathrm{global}$, plotted against (left) gravity $g$, and (right) the global compressibility,
$\kappa=mg/(k_n\langle d\rangle)$, on a log-linear scale for different values of the normal stiffness as shown in the legend.
The solid line represent \eref{eq:mu-kappa}.}
\label{fig:mu_kappa}
\end{figure}

Figure \ref{fig:mu-g} displays the global friction coefficient $\mu_0^\mathrm{global}$ plotted against gravity $g$ for different values of the normal stiffness $k_n$,
 as given in the legend. We observe that $\mu_0^\mathrm{global}$ decreases with increasing gravity, while it increases with increasing $k_n$.
 Figure \ref{fig:mu_g_kappa} shows the global friction coefficient with different values of the normal stiffness $k_n$ and gravity $g$,
 where all results of $\mu_0^\mathrm{global}$ are collapsed if we plot them against $\kappa=\kappa^2_g$ ($\kappa_g$ is given by \eref{eq:kappa}).

In figure \ \ref{fig:mu_g_kappa}, the solid line is given by
\begin{equation}
\label{eq:mu-kappa}
\mu_0^\mathrm{global} = \mu_r^\mathrm{global} - \left(\frac{\kappa}{\kappa_0}\right)^\alpha~,
\end{equation}
where $\mu_r^\mathrm{global}=0.17\pm 0.01$ is the global friction coefficient in the rigid particle limit
and the exponent and characteristic global compressibility are given by $\alpha\simeq 0.5$ and $\kappa_0\simeq2.01$, respectively.

Previous microgravity parabolic flight and centrifuge experiments
\cite{macari1991analysis,white1990dynamic,alshibli2000constitutive,sture1998mechanics,brucks2007behavior,alshibli1996mechanics}
 showed a similar decreasing trend of the macroscopic friction coefficient on gravity. Few authors \cite{white1990dynamic,sture1998mechanics} attributed this dependence
 to the fact that at low gravity, the body forces become weak and the electrostatic cohesive forces begin to dominate.
 Klein \etal \cite{white1990dynamic} also argued that a load-dependent interparticle friction coefficient might be responsible for this behavior.
 However, no cohesive force or load-dependent friction was implemented in any of the DEM simulation data presented here.
 Hence, we claim that there should be an additional mechanism responsible for this interesting behavior.
In order to gain a better understanding in the following, we have a closer look by studying the system locally.

\subsubsection{Local compressibility}
\label{subsub:nonlinear}
Since our system is heterogeneous in both stress and strain rate fields, a local description of the system is
 highly desirable. In the shear stress-pressure ($\tau-p$) curves for different softness and gravity (figure\ \ref{fig:tau-p-k-g}),
the dependence of shear stress on pressure slightly ``bends" with increasing softness and gravity, i.e.,
 the friction coefficient has a dependence on the pressure and the shear stress becomes a nonlinear function of pressure as:
\begin{equation}
\tau(r,h) = \mu_0^\mathrm{local}(p,k_n) p(r,h)~,
\label{eq:local_mu}
\end{equation}
where $\mu_0^\mathrm{local}(p,k_n)$ is a \emph{local friction coefficient} which depends on pressure and contact stiffness.

\begin{figure}
\centering
\includegraphics[scale=0.5,angle=-90]{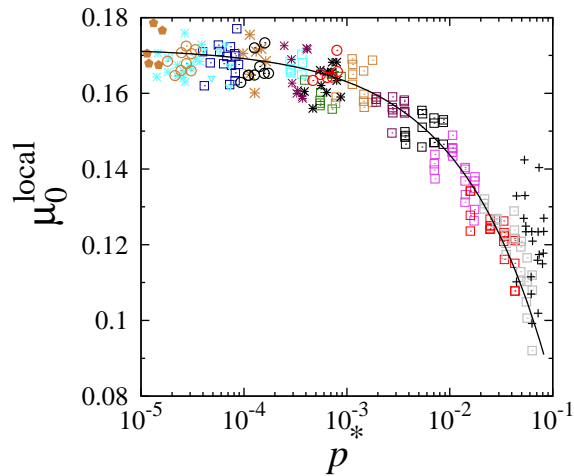}
\caption{(Color online)
The local friction coefficient, $\mu_0^\mathrm{local}(p^\ast)$, plotted against the local compressibility, $p^\ast$, on a log-linear scale.
Different symbols represent different values of $\kappa$ as given in the legend of figure\ \ref{fig:nu_P-kappa},
 while the solid line represents \eref{eq:mu-p}.}
\label{fig:kappa_sD}
\end{figure}

Figure \ref{fig:kappa_sD} shows the local friction coefficient with different values of the normal stiffness and gravity,
where all results of $\mu_0^\mathrm{local}(p,k_n)$ collapse if we introduce the \emph{local compressibility},
\begin{equation}
p^\ast\equiv\kappa^2_p=\frac{p\langle d\rangle}{k_n}~,
\label{eq:p_ast}
\end{equation}
defined as square of the ratio between two time scales, $T_k$ and $T_p$. $p^\ast$ can also be interpreted as non-dimensional average overlap (scaled with mean particle diameter).
In this figure, we scanned through a wide range of $p^*$ by systematically varying $g$ and $k_n$.
 We observe that for $p^* < 5\times 10^{-4}$, $\mu_0^\mathrm{local}(p^\ast)$ is almost constant,
 while for higher values, $\mu_0^\mathrm{local}(p^\ast)$ decreases with $p^*$ up to $p^\ast\approx 0.1$.

This dependence can be written in the form,
\begin{equation}
\mu_0^\mathrm{local}(p^\ast)=\mu_{\mathrm {r}}^\mathrm{local} - \left(\frac{p^*}{p^*_\sigma}\right)^{\beta_1},
\label{eq:mu-p}
\end{equation}
where $\mu_{\mathrm {r}}^\mathrm{local}=0.172\pm 0.01$ is the value of macroscopic friction in the rigid limit, which is in fair agreement with
  contact dynamics shear simulations for the same particle friction coefficient \cite{unger2010collective}.
The exponent is found to be $\beta_1 \approx 0.5\pm 0.04$ and the characteristic local compressibility is
 ${p^*_\sigma}=10.1 \pm 0.2$. As one extreme of $p^*$, for $p^*=0.1$ the average overlap is almost
 $10\%$ relative to the mean particle diameter, i.e., the \emph{soft particle limit}.
 The upper bound of $\mu_0^\mathrm{local}(p^\ast)$ is the low compression case, i.e., \emph{the rigid limit},
 where the average overlap is much smaller (1/10000) compared to the particle diameter
and $\mu_0^\mathrm{local}(p^\ast)$ is almost double as large as for $p^*\approx 0.1$.

\subsection{Local volume fraction}
\label{sub:volu}
\begin{figure}
\centering
\includegraphics[scale=0.5,angle=-90]{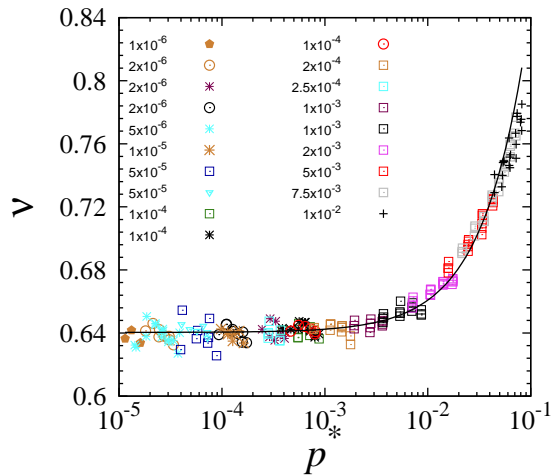}
\caption{(Color online)
The local volume fraction, $\nu(r,h)$, in the system plotted against the local compressibility, $p^*$, on a log-linear scale.
Different symbols represent different values of $\kappa$ as given in the legend.
The solid line represents \eref{eq:nu-P}.}
\label{fig:nu_P-kappa}
\end{figure}

In figure\ \ref{fig:nu_P-kappa}, the local volume fraction $\nu(r,h)$ is plotted against the local compressibility, $p^*$.
Because of slow quasi-static flows, no strong dilation is observed, i.e., no strong dependence of $\nu$ on local shear rate is observed. 
 We observe that the packing is rather loose for lower $p^*$ and tends to a critical value $\nu_c=0.642 \pm 0.002$, as observed in \cite{dorbolo2011influence}.
 The data can be fitted well by the functional form
\begin{equation}
p^*=a^* \left( \nu-\nu_{\rm {c}} \right),
\label{eq:nu-P}
\end{equation}
with $a^*=0.48\pm 0.02$ ($a^*$ can be further expressed in terms of volumetric fabric as reported in \cite{imole2012hydrostatic,goncu2010constitutive}).
Interestingly, no significant difference in volume fraction $\nu$ is observed for $p^*<10^{-3}$,
while for $p^*>10^{-3}$ within the fluctuations, $\nu$ increases almost linearly with $p^*$ (the curvature is due to the logarithmic $p^*$ axis).
The relation between $\nu$ and $p^*$ is well established in the case of static packings \cite{imole2012hydrostatic,goncu2010constitutive,zhang2005jamming}.
Here we show that the same relation holds for a slow granular flow, involving considerable finite but small strain rates.

\subsection{Local structure}
\label{sub:micor}
Shearing of a granular assembly always leads to the buildup of a contact anisotropy in the system \cite{veje1999kinematics,majmudar2005contact,azema2014internal}.
To study contact anisotropy in our system, we analyze the deviatoric fabric as defined in \eref{eq:Fabric}.
 We use the second invariant or $F_{\rm dev}$ \eref{eq:qdev} of the deviatoric fabric tensor to quantify anisotropy of the contact network in the system.

\subsubsection{Anisotropy}
\label{subsub:anis}
\begin{figure}
\centering
\includegraphics[scale=0.5,angle=-90]{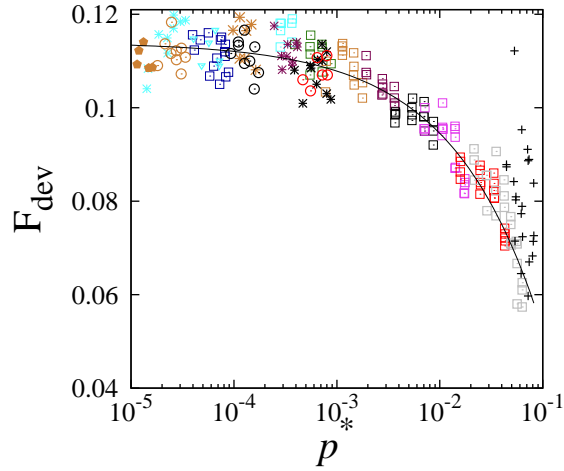}
\caption{(Color online)
The local deviatoric fabric, $F_{\rm dev}(r,h)$, plotted against the local compressibility, $p^*$, on a log-linear scale.
Different symbols represent different values of $\kappa$ as given in the legend of figure\ \ref{fig:nu_P-kappa}.
The solid line represents the corresponding fit to \eref{eq:FD-p}.}
\label{fig:kappa_FD}
\end{figure}

\
Figure \ref{fig:kappa_FD} displays the local deviatoric fabric, $F_\mathrm{dev}(r,h)$, plotted against the local compressibility $p^\ast$,
where $F_\mathrm{dev}(r,h)$ for different values of the particle stiffness and gravity is found to collapse on a unique curve (solid line).
This dependence can be written in a similar fashion to \eref{eq:mu-p},
\begin{equation}
F_{\rm dev}(p^\ast)=F^r_{\rm dev}-\left(\frac{p^\ast}{p^\ast_F}\right)^{\beta_2},
\label{eq:FD-p}
\end{equation}
where $F^r_{\rm dev}$ is the anisotropy of contact network in the rigid limit, the exponent is found to be $\beta_2 \approx 0.5 \pm 0.03$, and $p^\ast_F \approx 26.3 \pm 0.6$.
The decrease in $F_{\rm dev}$ with increasing $p^\ast$ can be explained in terms of the increasing volume fraction $\nu(r,h)$ with increase in $p^\ast$.
 In the case of a denser packing, particles have less free space to re-arrange (and thus align along preferential direction) and build up anisotropy in response to the local shear,
 compared to a relatively loose packing (at low $p^\ast$).
% This results in decrease in anisotropy of the contact network with increasing $p^\ast$.

%

\begin{figure}
\centering
\includegraphics[scale=0.5,angle=-90]{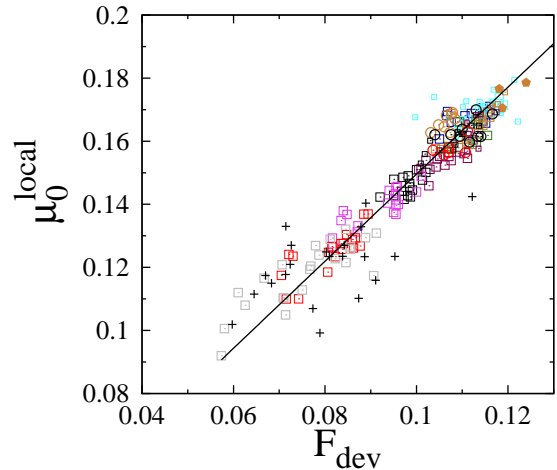}
\caption{(Color online)
The local friction coefficient, $\mu_0^\mathrm{local}(p^\ast)$, plotted against the local deviatoric fabric, $F_{\rm dev}(r,h)$, for different values of $\kappa$.
Different symbols represent different values of $\kappa$ as given in the legend of figure\ \ref{fig:nu_P-kappa}.
The solid line represents the corresponding fit to \eref{eq:mu_local-F_dev}.}
\label{fig:sD-FD_norm-kappa}
\end{figure}

In figures \ref{fig:kappa_sD} and \ref{fig:kappa_FD},
we observe that the the local shear resistance and the local anisotropy of the contact network in the quasi-static state show a similar trend.
In figure\ \ref{fig:sD-FD_norm-kappa}, we plot $\mu_0^\mathrm{local}(p^\ast)$ against $F_{\rm dev}(p^\ast)$ for different values of $\kappa$, where
 a linear correlation can be inferred as,
\begin{equation}
\mu_0^\mathrm{local}(p^\ast)=\mu_{\rm {iso}}+bF_{\rm dev}(p^\ast)~,
\label{eq:mu_local-F_dev}
\end{equation}
where $\mu_{\rm {iso}}=0.01\pm 0.01 (\approx 0)$ is the friction coefficient in the (extrapolated) limit of the {\em isotropic} contact network $(F_{\rm dev}=0)$
 and $b=1.38\pm 0.02$ is a constant of proportionality.
 This clearly shows that the shear resistance seems to accompany the anisotropy in the contact network in the critical state.
 It is worthwhile to mention that no information about such a relation in the transient regime can be inferred from our analysis,
 while it is strongly supported by our data in the critical state. 
 It also links the increase in the friction coefficient with decreasing gravity
 (as observed in figure\ \ref{fig:mu-g}) to the change in the contact network configuration of the material at lower $g$.

\subsubsection{Shape factor}
\label{subsub:shap}
In this section, we compare the shape factor ($\frac{Q_2}{Q_1}$) for stress and fabric tensors, where $Q_2$, and $Q_1$ are the eigenvalues
 of the deviatoric tensors as defined in section \ref{sec:tensor-iso-dev}.
 The shape factor quantifies the out of shear plane neutral eigenvalue, compared to the maximum eigenvalue.
 Note that for strain rate tensor $\dot{\epsilon}_2/\dot{\epsilon}_1\equiv0$ due to geometry and symmetry, i.e., we have plain-strain shear.

\begin{figure}
\subfigure{\includegraphics[scale=0.5,angle=-90]{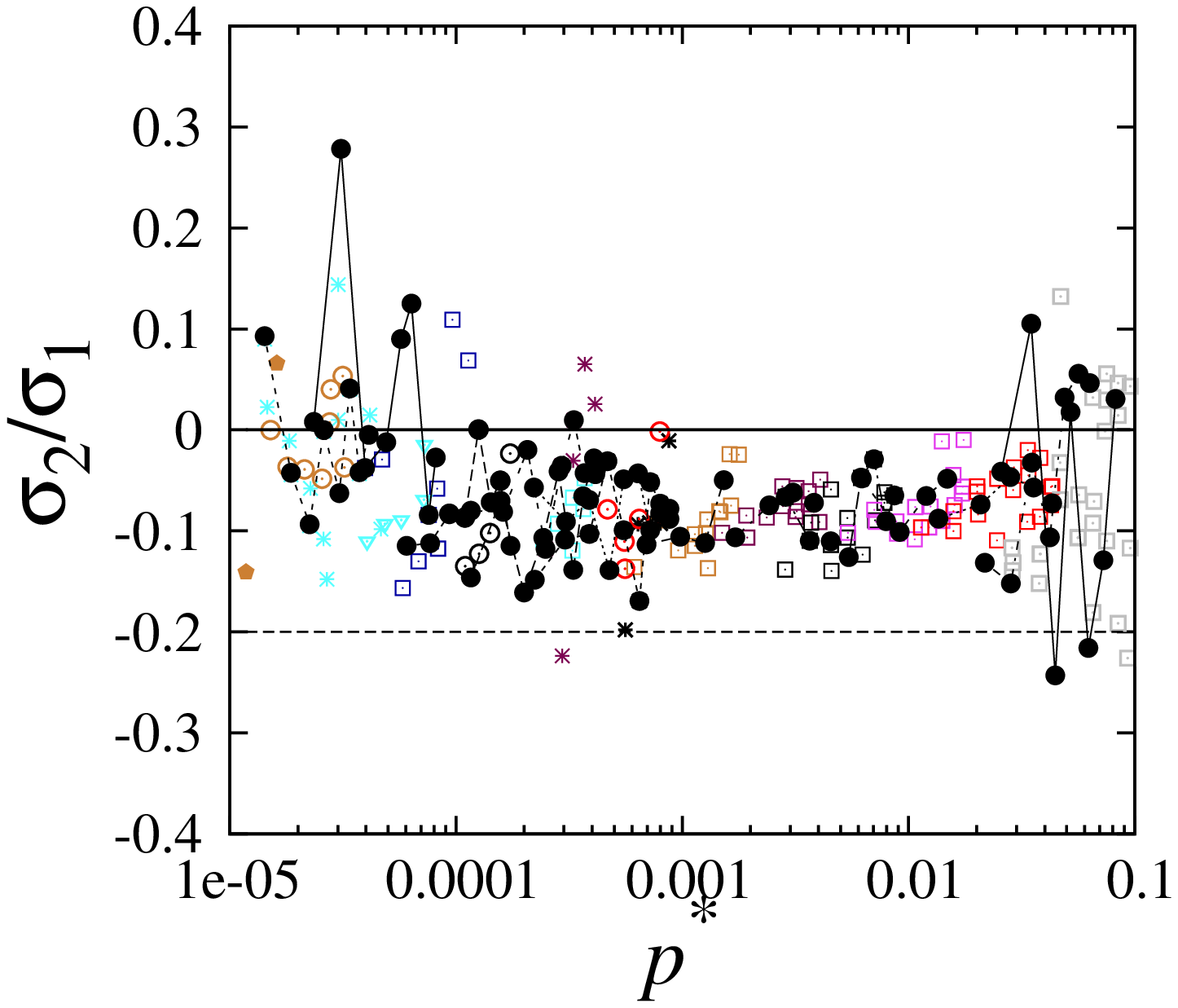}\label{fig:sigma-comp}}
\subfigure{\includegraphics[scale=0.5,angle=-90]{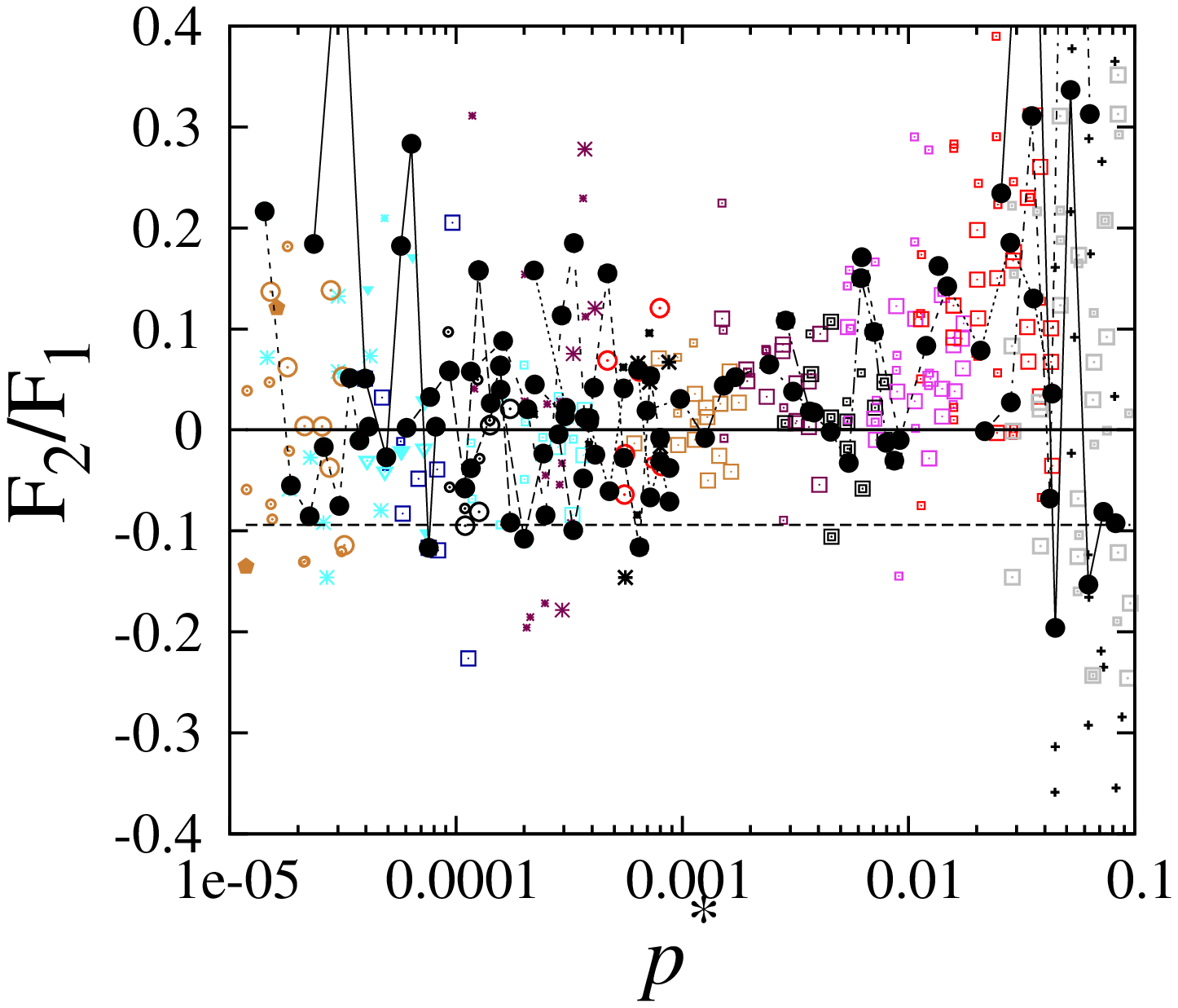}\label{fig:fab-comp}}
\caption{(Color online)
Shape factor for (left) stress, and (right) fabric plotted against dimensionless pressure $p^*$.
Different symbols represent different values of $\kappa$ as given in the legend of figure\ \ref{fig:nu_P-kappa}.
 Black circles represent the data in the center of the shear band. Solid line represents zero, while dashed line is the prediction
 from \cite{weinhart2013coarse}.}
\label{fig:shape-comp}
\end{figure}

In figure\ \ref{fig:sigma-comp}, we plot the shape factor for the stress tensor. Different symbols represent different values of compressibility $\kappa$ as given in the legend of figure\ \ref{fig:nu_P-kappa},
 while black circles show the data in the center of shear bands for these simulations.
 We observe that the shape factor is highly fluctuating for the two extremes of $p^*$, while
 in the range $10^{-4}<p^*<10^{-2}$, it is significantly below zero. This implies that the stress in the shear plane is higher as compared to the axial stress along the neutral direction orthogonal to the shear plane.
 The sign means that this axial stress is reduced perpendicular $-1$ and enhanced $+1/2$ within the shear plane \cite{weinhart2013coarse}. The dashed line is the data from
 \cite{weinhart2013coarse}, where authors studied the flow behavior on an inclined plane. We observe that the sign for both the cases is negative, while
 the magnitude is different, which might be due to the difference in interparticle friction. In figure\ \ref{fig:fab-comp}, the shape factor of the fabric tensor fluctuates (strongly) around zero.
 It is important to mention that the fluctuations in the data are from a single simulation.
%Negative shape factor implies the expansion of the system. 

These two observations suggest that the fabric and stress tensors behave differently even though they are proportional in magnitude (norm) as shown in figure\ \ref{fig:sD-FD_norm-kappa}.
The fabric tensor is in a planar state like the strain rate tensor, whereas the induced stress state is triaxial, as expected for a solid-like material \cite{weinhart2013coarse}.
$F_2/F_1$ tends to positive values for larger $p^*$, further establishing the difference between structure and stress tensors. However, in order to have a clear picture
 for the fabric tensor, the strong and weak subnetworks should be studied separately, since only the strong subnetwork carries almost all of the fabric anisotropy \cite{imole2014micro,singh2014effectf}.
%

%% Comment about lowering the cutoff
As discussed in section \ref{sub:stress-sb} the cutoff shear rate $\dot{\gamma}_c$ can depend on the simulation time or the averaging time. In this section, we focused
on the data only inside the shear band, which are in the critical state and have forgotten their initial configuration due to large strain.
 However, the velocity gradients in the setup are smooth, which implies
 that part of the system outside the shear band is also flowing, but only very slowly. If the simulation runs longer, the cutoff can be lowered,
 eventually if the simulation would run infinitely long,
 no cutoff on the local strain rate is needed.
If we lower the cutoff on local strain rate (see next section), by setting $\dot{\gamma}_c(\Omega)\equiv \frac{\Omega}{2\pi}$, we observe 
much wider variation of data in our results, especially for the local friction coefficient, the deviatoric fabric and the volume fraction.
 However, the shape factors are not
 strongly influenced by a change in $\dot{\gamma}_c$, by nearly an order of magnitude.

\section{Complete rheology}
\label{sec:iner}
The previous section showed that in the quasi-static state the friction coefficient and deviatoric fabric are strongly correlated
 in the critical state. Motivated by this, we check if this correlation also works in the rate-dependent inertial regime.
 To test the correlation, high inertial number data are generated by varying the external rotation rate $\Omega$ for a fixed gravity and contact stiffness.
 In the following, we will explore the evolution of the local macroscopic friction coefficient $\mu$ and deviatoric fabric
 with $I$ and combine it with the dependence of both $\mu$ and deviatoric fabric on $p^*$, to propose the complete rheological law.
 It is important to mention that in this section the cutoff on local strain rate is lowered to $\dot{\gamma}_c(\Omega)\equiv \frac{\Omega}{2\pi}$, so as to capture the maximum data and present a unique rheology law outside and inside the shear band.

\subsection{Friction law}
\label{sub:iner_fric}
\begin{figure}
\centering
\includegraphics[scale=0.5,angle=-90]{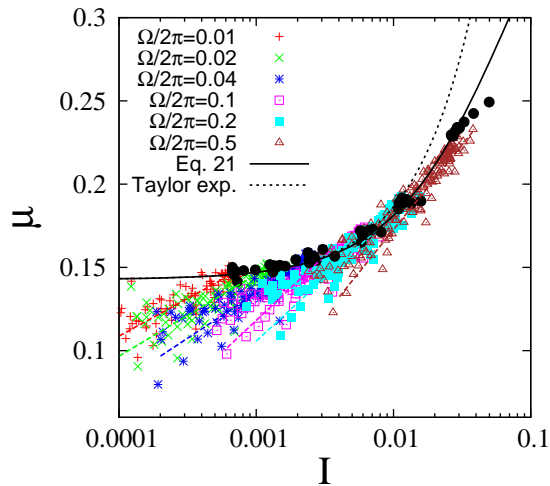}
\caption{(Color online)
The friction coefficient plotted against the inertial number $I$ for results 
from simulations with different rates of rotation.
Different symbols represent different rates of rotation as given in the legend, lines with the same color
 represent solution of \eref{eq:koval}  with $\mu_0^{\mathrm{local}}=0.14$, $\mu_2^{\mathrm{local}}=0.3$ and $I^{\sigma}_0=0.026$.
Dot-dashed line shows the Taylor expansion of \eref{eq:koval}.
 Black circles represent the data in the center of the shear band.
}
\label{fig:sD-I-koval}
\end{figure}

The local friction coefficient $\mu$ is plotted against inertial number $I$ in figure\ \ref{fig:sD-I-koval}.
 Different symbols show data from
 different rates of rotation as given in the inset; the black solid circles represent the data in the
 center of the shear band. The solid black line shows the friction law, as proposed in \cite{JopForterrePouliquen2006}:

\begin{equation}
\mu(I,p^*) = \mu_0^{\mathrm{local}}(p^*) + \frac{\mu_2^{\mathrm{local}}(p^*)-\mu_0^{\mathrm{local}}(p^*)}{1+I^{\sigma}_0/I}~,
\label{eq:general}
\end{equation}
 with $\mu_0^{\mathrm{local}}(p^*)$, as given in \eref{eq:mu-p}. We observe that data from simulation using
 a single gravity and contact stiffness does not give a wide variation in $\mu$ and $\mu_0^{\mathrm{local}}=0.14$, $\mu_2^{\mathrm{local}}=0.5$ and $I^{\sigma}_0=0.1$
 fits well the data.
 A Taylor expansion (in the range $I< I^{\sigma}_0$ )
 to above equation is $\mu(I) \cong \mu_0^{\mathrm{local}} + ({\mu_2^{\mathrm{local}}-\mu_0^{\mathrm{local}}})\frac{I}{I^{\sigma}_0}$, which is
 similar to the frictional law proposed in \cite{midi2004dense,da2005rheophysics}.
 Two different trends emerge, i.e., the shear band center data can be very well fitted by \eref{eq:general} and
 for $I\ge0.01$ data collapse on a unique curve. On the other hand, for lower values of $I$, deviations
 from this relation are observed, depending on the external rotation rate. The friction coefficient in slow flows (steady state) becomes smaller
 than $\mu_0^{\mathrm{local}}$, i.e. in our system the granular material is able to flow below $\mu_0^{\mathrm{local}}$.
 The deviation of our data from the main law \eref{eq:general} is consistent with observations in \cite{KovalRouxCorfdirChevoir2009,KamrinKoval2012},
 where this deviation was explained based on the heterogeneity in the stress field (due to strain rate). In our system, we have gradients in stress arising due to
 gradients in both strain rate and pressure (due to gravity).

In order to quantify the deviation from \eref{eq:general}, we fit the data with:
\begin{equation}
\mu(I<I^*,p^*)=\mu_0^{\mathrm{local}}(p^*)\left[ 1-\alpha \ln{(I/I^*)}\right]  ~,
\label{eq:koval}
\end{equation}
 where $\alpha$ is a constant and $I^*$ is the characteristic inertial number when $\mu\cong \mu_0^{\mathrm{local}}$.
 This relation is similar as proposed in \cite{KovalRouxCorfdirChevoir2009} for two-dimensional ring shear cell data.
 As the above relation was derived for a two-dimensional setup with constant pressure, we fit it to our data at three
 different heights (i.e. pressure levels), close to top, at mid-height and close to bottom. In figure\ \ref{fig:sD-I-koval}, different colored dashed lines represent this fit at the mid-height of the system. We observe that the prediction is in close agreement
 with the data, even though our system has different dimension and boundary conditions.
 \ref{sec:mu_press1} shows the data and corresponding fits for different heights (pressures). We find that both $\alpha$
 and $I^*$ do not show a clear dependence on pressure. 

\subsection{Fabric anisotropy}
\label{sub:dense_aniso}
\begin{figure}
\centering
\includegraphics[scale=0.5,angle=-90]{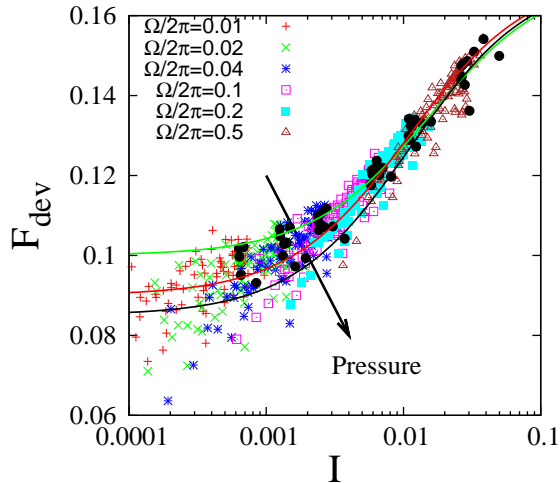}
\caption{(Color online)
The local fabric anisotropy $F_{\rm dev}$ plotted against the inertial number $I$ for results 
from simulations with different rates of rotation.
Different symbols represent different rates of rotation as given in the legend.
 Black circles represent the data in the center of the shear band. Green, red, and black
 lines are fit to \eref{eq:general-Fdev} for  pressure levels $p=100, 200,$ and $400$\ $\rm{Nm^{-2}}$ respectively,
 with fit parameters given in Table \ref{tab:fitparameters}. The arrow shows increasing pressure.
}
\label{fig:Fdev-I}
\end{figure}

In order to look for the connection between anisotropic fabric and macroscopic friction coefficient in the inertial regime,
 here we explore the dependence of $F_{\rm dev}$ on $I$.
In figure\ \ref{fig:Fdev-I}, we plot the local $F_{\rm dev}$ as obtained by simulations with different rates of rotation against $I.$ 
 We observe that like $\mu$, $F_{\rm dev}$ varies strongly against $I$ and its dependence on $I$ can be represented as:
 \begin{equation}
F_{\rm dev}(I,p^*) = F_{\rm dev}^0(p^*) + \frac{F_{\rm dev}^{(2)}(p^*)-F_{\rm dev}^0(p^*)}{1+I^{F}_0/I}~,
\label{eq:general-Fdev}
\end{equation}
 with $F_{\rm dev}^0$ being the fabric anisotropy in the quasi-static state, $F_{\rm dev}^{(2)}$ is the limiting fabric anisotropy,
 and $I^{F}_0$ is the typical inertial number, which is an order of magnitude different from $I^{\sigma}_0$. Green, red and black lines show the fit to above relation at pressure levels
 100, 200 and 400 $\rm{Nm^{-2}}$, respectively, with points inside the shear band highlighted (black circles).
 Fit parameters to these results are presented in table \ref{tab:fitparameters}. It is noticeable that unlike $\mu$, $I$ alone 
 is not able to describe $F_{\rm dev}$, with the effect of pressure being prominent in case of slow flows i.e., low $I$.
 In contrast, for fast flows, the anisotropy seems to become independent of pressure.

 The increase in the contact anisotropy with inertial number is in accordance with the recent studies \cite{weinhart2013coarse,azema2014internal}.
 It is important to mention that for even higher rates of rotation of the system, i.e. inertial number $I>0.1$, $F_{\rm dev}$ shows
 a different behavior as predicted by \eref{eq:general-Fdev}
 and a decreasing trend is observed (as reported in \cite{phdsinghmicro}), which is beyond the scope of this work.
 This might be due to the fact that for $I>0.2$ the packing becomes very loose ($\nu\le0.55$).
 Also for such high rates of rotation, the centrifugal force on grains due to rotation becomes
 comparable to the gravitational force. As a result, the top surface is not flat anymore, instead the surface develops
 a dip in the middle, as observed previously \cite{phdsinghmicro,dijksman2009granular,corwin2008granular}. In this range,
 also the kinetic and contact contributions of the local macroscopic friction $\mu$ become comparable. 

%
%%%%%%%%%%%%%%%%%%%%%%%%%%%%%%%%%%%%%%%%%%%%%%%%%%%%%%%%%%%%%%%%%%%%%%%%%%%%%%%%%%%%%%%%%%%%%%%%%%%%%%%%%%%%
\begin{table*}
\centering
\begin{tabular}{|l|l|l|l|}
\hline
    $p$ & $F_{\rm dev}^0$ & $F_{\rm dev}^{(2)}$ & $I^{F}_0$ \\
\hline
    100 & 0.1$\pm$ 0.0005 & 0.17$\pm$ 0.0005 & 0.012 \\
    200 & 0.095$\pm$ 0.0008 & 0.17$\pm$ 0.0001 & 0.011 \\
    400 & 0.085$\pm$ 0.0001 & 0.17$\pm$ 0.0004 & 0.009 \\
\hline
\end{tabular}
\caption{Table showing the fit parameters $F_{\rm dev}^0$, $F_{\rm dev}^{(2)}$, and $I^{F}_0$ in \eref{eq:general-Fdev}
 for different values of pressure $p$ (in units of ${\rm Nm^{-2}}$).}
\label{tab:fitparameters}
\end{table*}
%%%%%%%%%%%%%%%%%%%%%%%%%%%%%%%%%%%%%%%%%%%%%%%%%%%%%%%%%%%%%%%%%%%%%%%%%%%%%%%%%%%%%%%%%%%%%%%%%%%%%%%%

\begin{figure}
\centering
\includegraphics[scale=0.5,angle=-90]{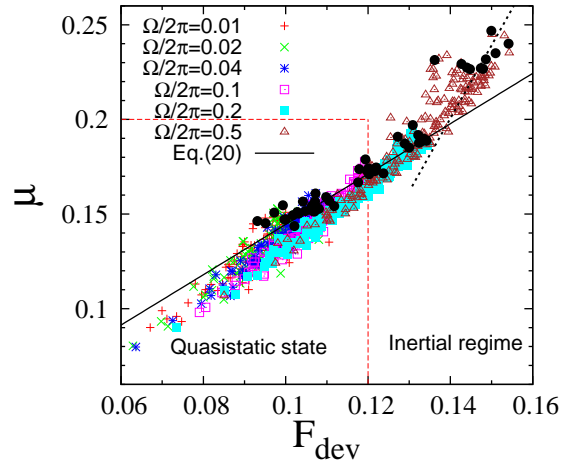}
\caption{(Color online) $\mu$ plotted against $F_{\rm dev}$ for data in previous section and different rates of rotation.
Different symbols represent different rates of rotation as given in the legend.
The solid line represent \eref{eq:mu_local-F_dev}, while the dashed line (with slope $\backsimeq 3.5$) is fit by the eye.
Black circles represent the data in the center of the shear band.}
\label{fig:FD_sD_full}
\end{figure}

Starting from both variations of macroscopic friction and fabric anisotropy as a function of inertial number $I$,
 it is tempting to ask the question if the correlation in \eref{eq:mu_local-F_dev} holds for the inertial regime as well.
 The result is displayed in figure\ \ref{fig:FD_sD_full}. Solid line shows \eref{eq:mu_local-F_dev}, which fits well the shear band center data
 being highlighted by black circles. Here again, we find data outside the shear band showing a different trend. Red dashed line separates
 the data into quasi-static and dense inertial regimes. It is interesting to note that \eref{eq:mu_local-F_dev} (as shown by black solid line)
  very well captures the trend at the onset of dense inertial regime. However, for faster flows, where rate dependence becomes prominent, a different trend
 is observed which can also be fitted by a slightly different linear relation (dot-dashed line). This shows that the macroscopic friction
 and fabric anisotropy are correlated even in the dense inertial regime.

\section{Discussion and Conclusion}
\label{sec:conc}
 To summarize, this work is an exploration of the flow behavior for 3D granular shear flow using DEM simulations. We particularly focused on the effect
 of external compression (gravity) and the particle softness on the flow behavior, considering both stress and structure.

\paragraph{Quasi-static flows}
 Our study shows that the shear strength (macroscopic friction coefficient $\mu$) of the material decreases with increase in either gravity or particle softness for the quasi-static flows.
 We find that the data for different gravity and particle softness can be expressed as a unique power law, when analyzed in terms of a control parameter called
 {\em local compressibility} $p^\ast$ (as defined in \eref{eq:p_ast}). $p^\ast$ can be interpreted as the non-dimensional {\em local}
 average overlap (scaled with mean particle diameter) and can be used to quantify the softness of the bulk material relative to the {\em local} compression level. Low values of $p^\ast$ signify rigid particles, while a high $p^\ast$ implies soft, deformable particles.
 Both softness and gravity are also found to affect the local microstructure, i.e., the anisotropy of the contact network, which is quantified by the deviatoric fabric ($F_{\rm dev}$).
 We show that the deviatoric fabric can also be expressed as a power law of $p^\ast$ with an exponent similar to that of the shear strength.
 This points out that the local anisotropy of the contact network (deviatoric fabric) and the shear strength of the material are highly correlated
 in the slow,  quasi-static flows and the shear strength follows the anisotropy of the contact network. 

These results are highly significant for planetary studies regarding the shear strength of the granular material in extraterrestrial bodies such as Moon or Mars. Significant amount of experimental work using parabolic flight have shown the increase in shear strength  of the material with decreasing gravity
 \cite{macari1991analysis,costes1987microgravity,white1990dynamic,alshibli2000constitutive,alshibli1996mechanics,sture1998mechanics} without proper explanation.
 We propose that the anisotropy, i.e., the rearrangement in the contact network, is the key parameter for this dependence if no new forces are involved.
 With decreasing gravity, the packing becomes loose (due to decrease in body force acting on the particles), which in turn provides more free space to the particles to rearrange (and thus align)
 in response to the local shear. The fact that the particle softness and gravity have been shown to have similar effects on the local flow behavior,
 makes this work equally relevant for {\em soft particles}, that find their
 applications in many engineering and biological systems \cite{hoffman2002hydrogels}. 

Since it is extremely difficult and expensive to perform {\em in situ} experiments on the Moon (or the parabolic flight), the \textquoteleft compensation\textquoteright\
 effect we find with the ratio of gravity $g$ and particle stiffness $k_n$ allows to mimic variable gravity by tweaking/tuning the particle stiffness.

\paragraph{Inertial flows}
 Further, to study the rheology of the system for gravity $g=10\ \rm{ms^{-2}}$, the rotation rate of the system is increased. For faster flows the system enters into
 a rate dependent {\em inertial regime}, consistent with previous studies \cite{midi2004dense,da2005rheophysics,forterre2008flows}.
 We find that both the macroscopic friction $\mu$ and contact anisotropy show a similar increasing behavior with inertial number $I$.
 This shows that macroscopic friction and contact anisotropy are correlated also in the dense inertial regime. The increase of $\mu$ with $I$ as observed
 in the {\em inertial regime} is accompanied by the evolution of the microstructure (contact anisotropy) with increasing inertial number.
 This picture is consistent with the recent study of Az{\'e}ma \etal \ \cite{azema2014internal}.

\paragraph{Open issues}
 We find that  the frictional laws obtained from homogeneous shear flows \cite{da2005rheophysics} can be applied locally in the inertial regime,
 while they fail to predict the behavior of the material in the slower, quasi-static regime.
 The local rheology laws can be applied to our data in the center of the shear bands, where the strain-rate and stress gradients are zero,
 hence the material can be assumed to be homogeneously sheared. However, away from the center of the shear bands, in the quasi-static regime,
 we observe a nearly identical range of $\mu$ values corresponding to a completely different range of $I$. We find that in our system the material
 is able to flow even below $\mu_0^{\mathrm{local}}$, but only very slowly.
 We have shown that some observations
 can be explained by using an approach similar to Koval \etal \cite{KovalRouxCorfdirChevoir2009}.
 These deviations might be well captured using the non-local models by Kamrin and coworkers
 \cite{KamrinKoval2012,henann2013predictive,kamrin2014effect}; this work is in progress. Another related issue which remains untouched is the effect of particle softness and external compression (gravity here) on the non-locality.
 A study of effect of gravity on primary and secondary velocity fields, as done recently in \cite{mudroch2013gravity,mudroch2013convection} also deserve a further study. 

\paragraph{Conclusion}
The macroscopic friction (shear strength) of the material is found to be affected not only by the local shear rate, but also by external compression (gravity) and softness
 of the particles. While traditionally the inertial number, the ratio of
 stress to strain-rate time-scales, is dominating the flow rheology, we find that a second dimensionless number,
  the ratio of softness and stress time scales, must be involved to characterize the bulk flow behavior.
 For very slow shear rate the former can be ignored, while the latter two affect the shear strength by decreasing it with increase
 in either gravity (and thus local pressure) or particle softness. For faster flows, the macroscopic friction is found to increase in general with increasing shear rate.
 However, the tails of shear bands feature an anomalously small macroscopic friction - as observed previously \cite{KovalRouxCorfdirChevoir2009,KamrinKoval2012,kamrin2014effect}.
For the dependence of macroscopic friction on above three quantities, the  change in local microstructure (contact anisotropy) is found to be a
 key parameter, that was not often considered yet.

Looking towards the future, we are now in a position to address various important issues, such as unexpectedly high shear strength of the material
 at low (confining) stress or reduced gravity and a direct relation between the contact anisotropy and the  shear strength of the material. These issues
 are vital for a better explanation of the macroscopic behavior of the granular systems from a microscopic observation. The current study dealt with
 a dense system with small interparticle friction ($\mu_{\rm{p}}=0.01$), where the effect of softness on the macroscopic behavior is more direct than for large $\mu_{\rm{p}}$. However,
 an issue which remains unanswered and will be an extension of this study is whether the same effect can also be observed for relatively loose system
 (with higher interparticle friction). Further, the question whether the correlation between contact anisotropy and shear strength is just a consequence of relatively
 low interparticle friction or if it will also hold for a more realistic material (with higher interparticle friction) remains to be answered.  

\section{Acknowledgements}
We would like to thank C R K Windows-Yule for his careful revision. 
We thank W. Losert, D. van der Meer, M. Sperl, D. Lohse, B. Tighe, P. Jop, H. Hayakawa, A. Ikeda, O. I. Imole, and L. Silbert  for 
 stimulating discussions.
Financial support through the ``Jamming and
Rheology'' project of the Stichting voor 
Fundamenteel Onderzoek der Materie (FOM), 
which is financially supported by the 
``Nederlandse Organisatie voor Wetenschappelijk Onderzoek'' (NWO), 
is acknowledged. 

\appendix

\section{Pressure dependence of local macroscopic friction}\label{sec:mu_press1}
\begin{figure*}
\subfigure[]{\includegraphics[scale=0.33,angle=-90]{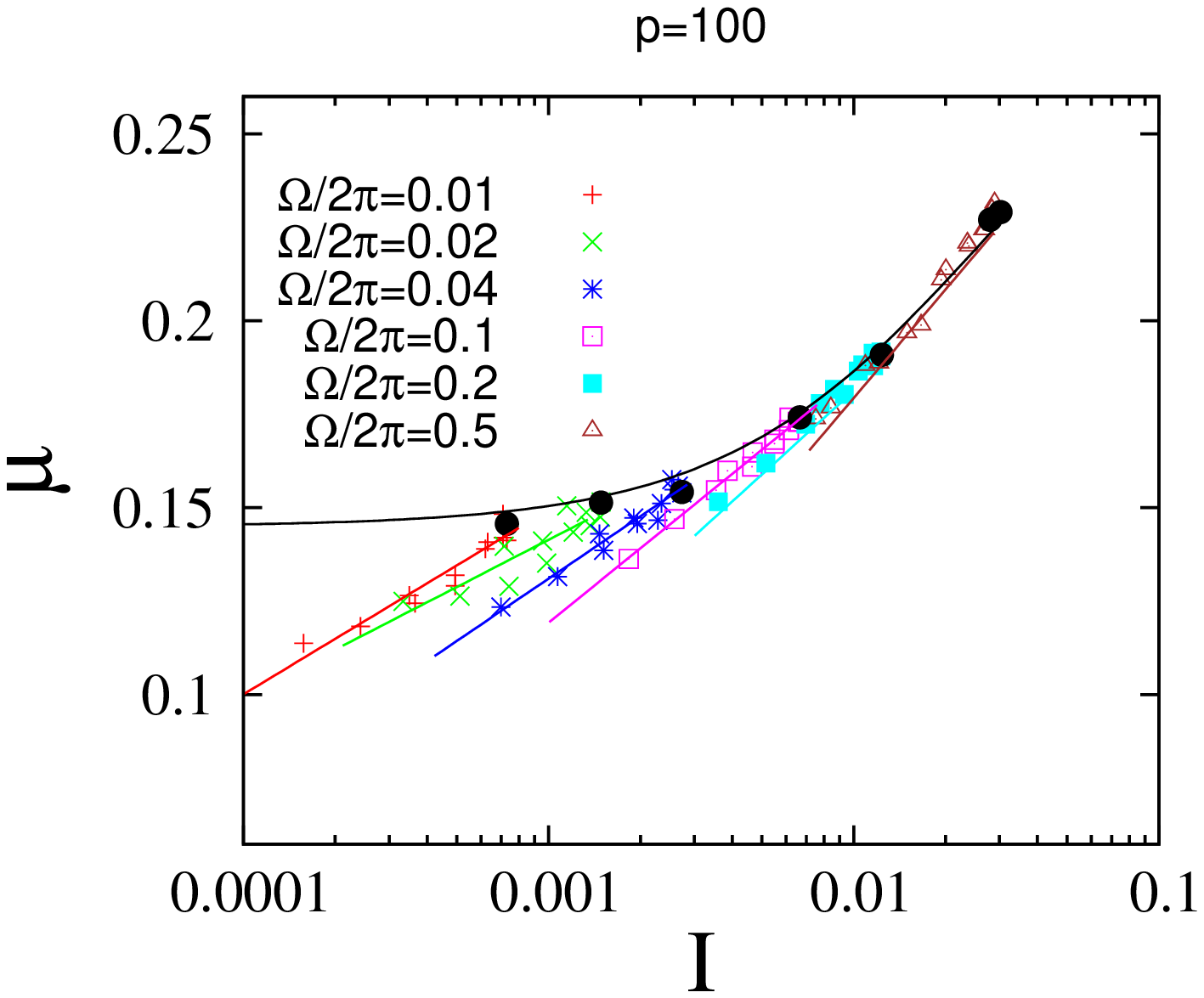}\label{fig:pos_top_coh1}}
\subfigure[]{\includegraphics[scale=0.33,angle=-90]{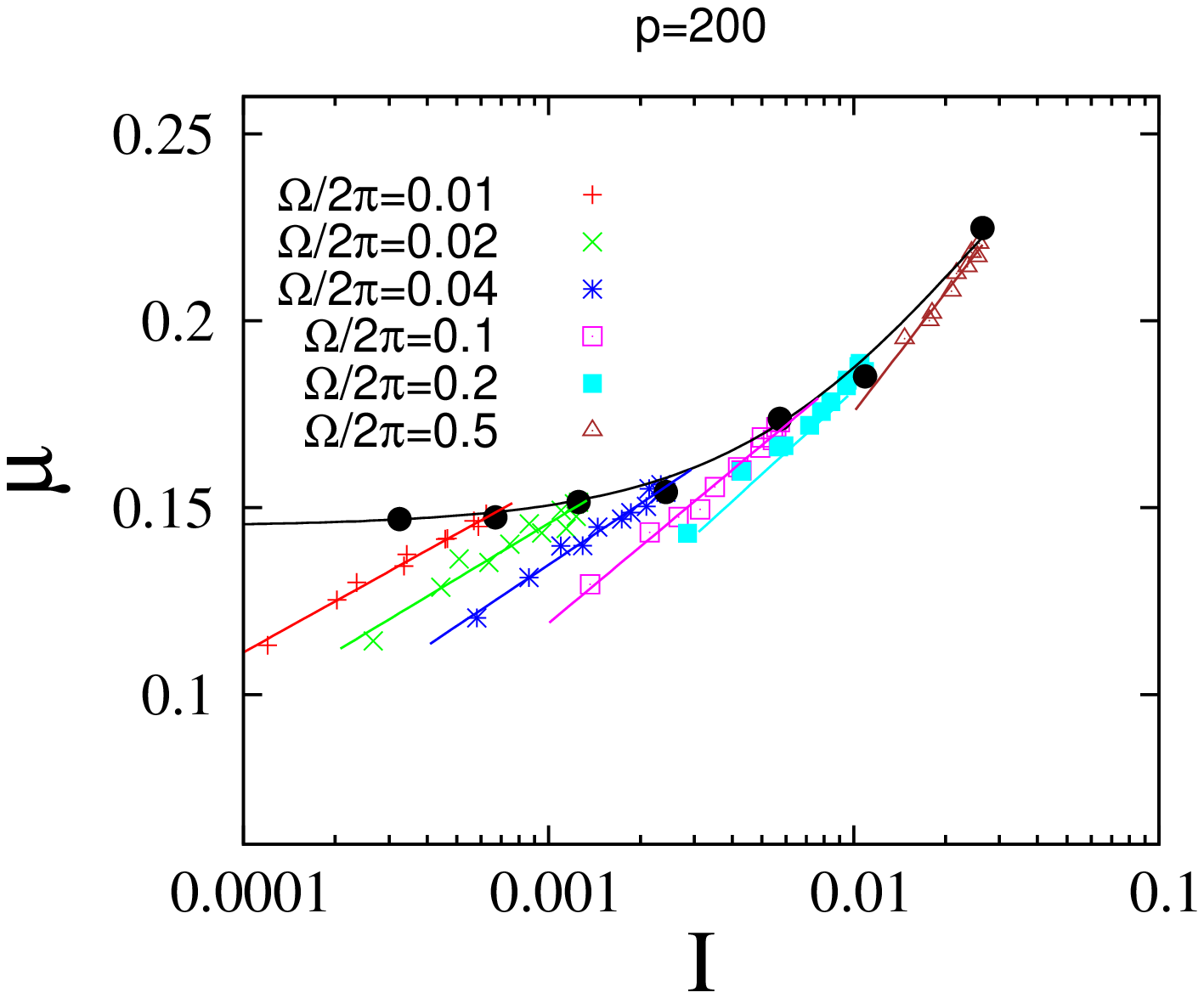}\label{fig:wid_top_coh2}}
\subfigure[]{\includegraphics[scale=0.33,angle=-90]{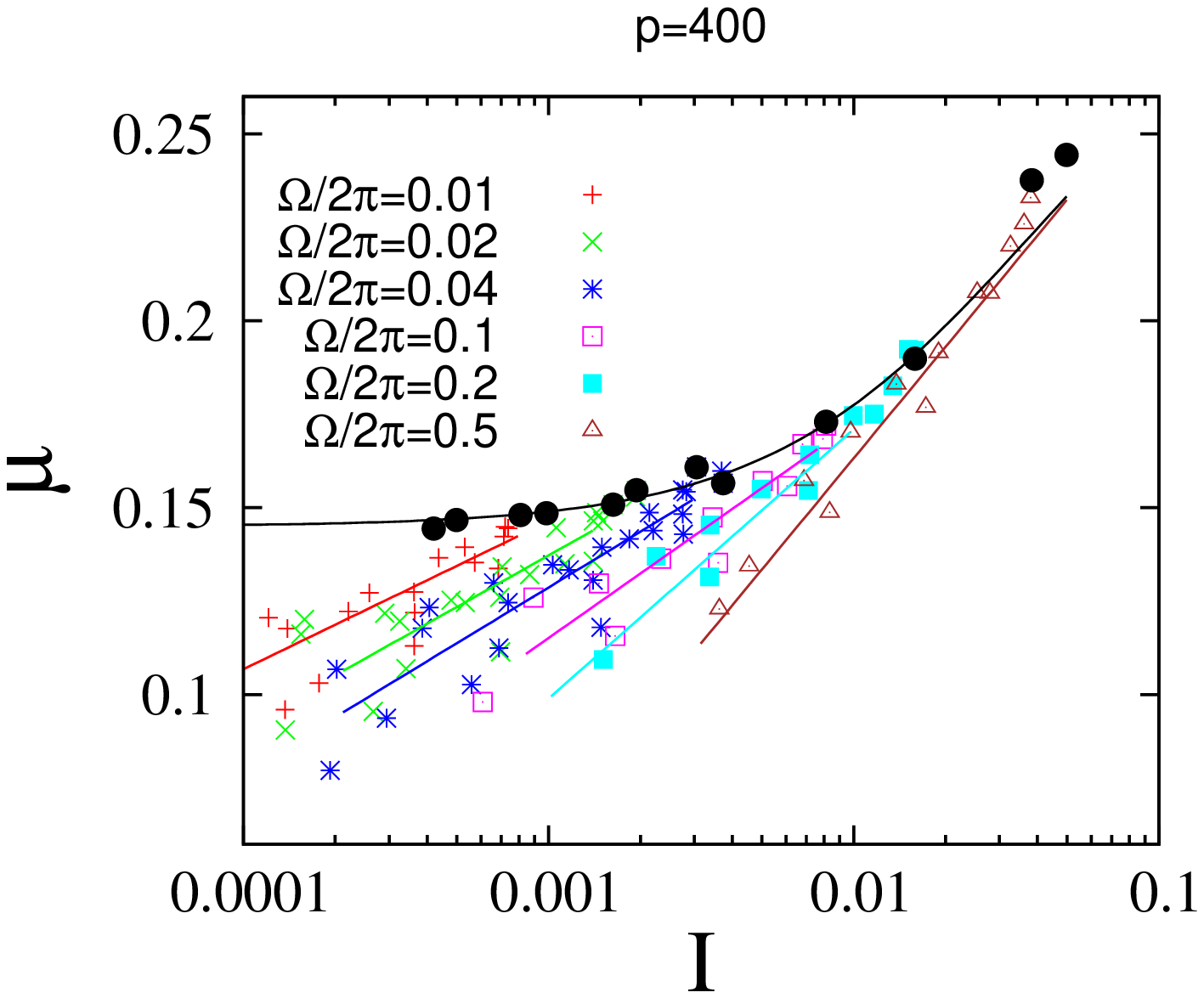}\label{fig:wid_top_coh3}}
\caption{(Color online)
$\mu$ plotted against $I$ for different local pressures in the system (a) $p=100$, (b) $p=200$, and (c) $p=400$ $\mathrm{Nm^{-2}}$.}
\label{fig:mu_press}
\end{figure*}

\begin{figure*}
\centering
\subfigure[]{\includegraphics[scale=0.33,angle=-90]{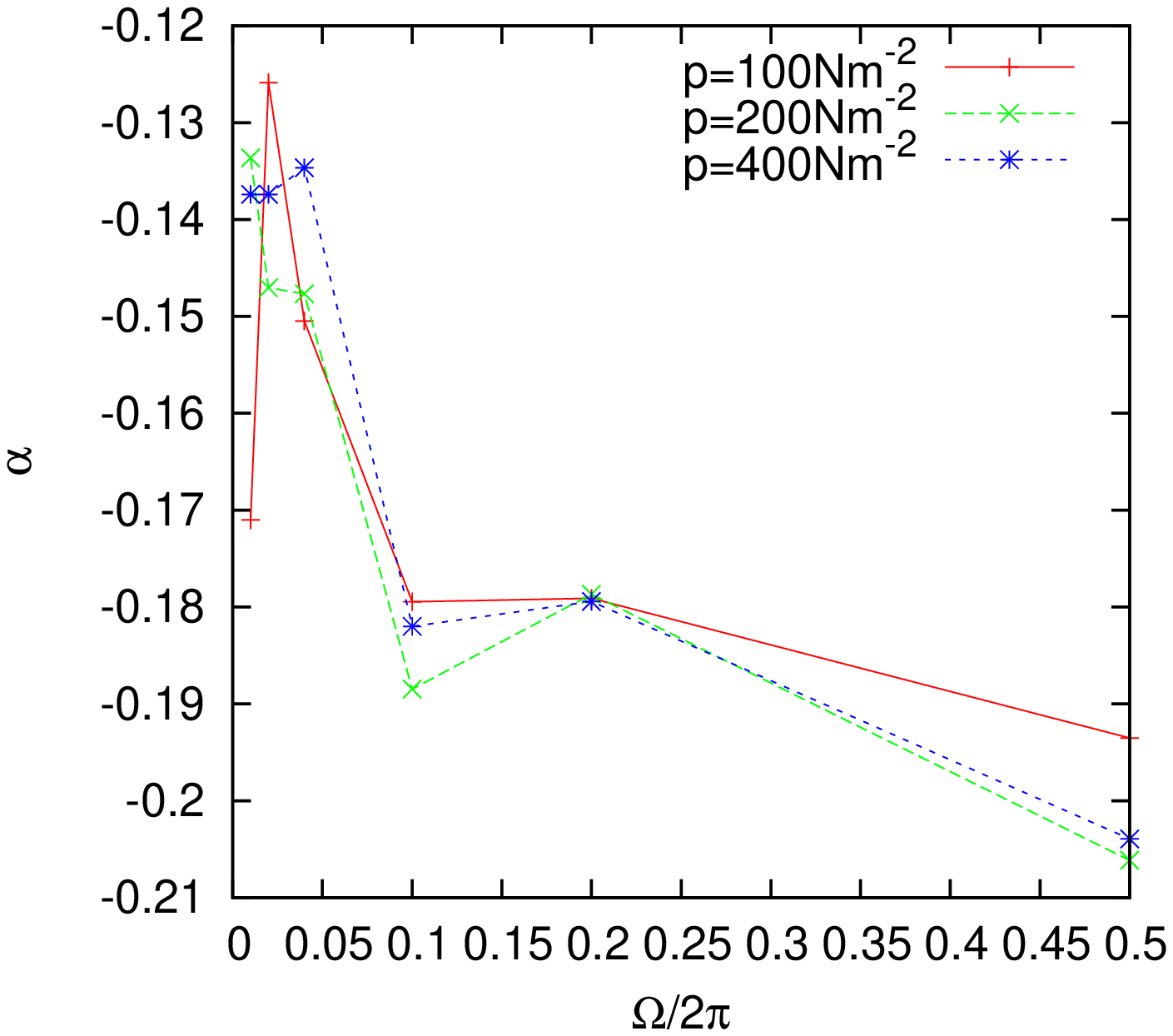}\label{fig:pos_top_coh}}
\subfigure[]{\includegraphics[scale=0.33,angle=-90]{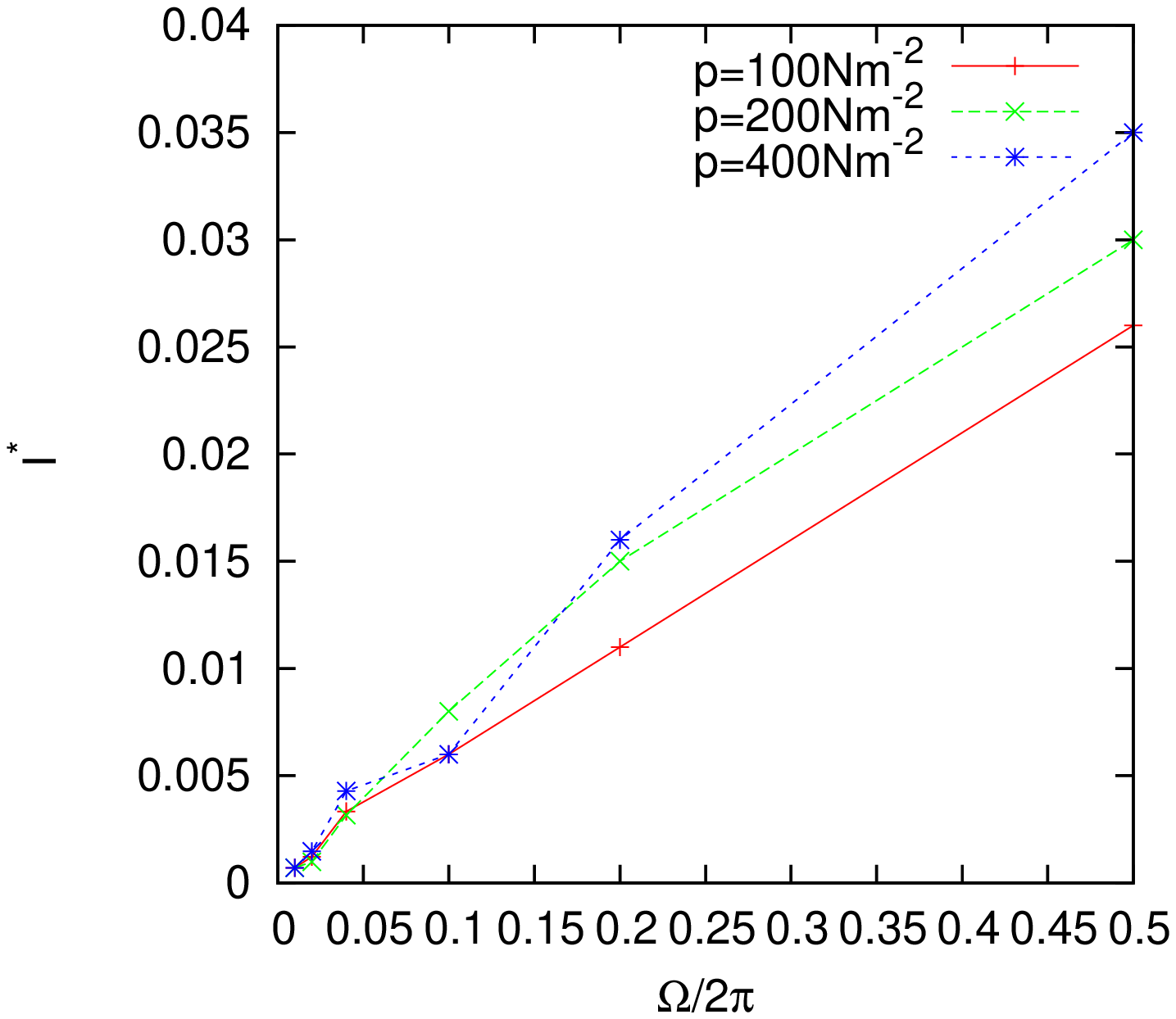}\label{fig:wid_top_coh1}}
\caption{(Color online)
(a) $\alpha$, and (b) $I^*$ plotted against the external rotation rate for different local pressures in the system.}
\label{fig:fit_press}
\end{figure*}

%%%%%%%%%%%%%%%%%%
In this section, we explore the pressure dependence of our rheological laws as presented in section \ref{sec:iner}.
Figure \ref{fig:mu_press} shows the fits for three different pressure levels (height) and we find for all of them the fitting
 is in good agreement with the data. Figure \ref{fig:fit_press} shows fit parameters for different pressure levels,
 where we observe that none of them show a clear strong dependence on pressure.
%%%%%%%%%%%%%%%%%%%%%%%%%%%%%%%%%%%%%%%%%%%%%%%%%%%%%%%%%%%%%%%%%%%%%%%%%%%%%%%%
%%%%%%%%%%%%%%%%%%%%%%%%%%%%%%%%%%%%%%%%%%%%%%%%%%%%%%%%%%%%%%%%%%%%%%%%%%%%%%%%%

%%%%%%%%%%%%%%%%%%%%%%%%%%%%%%%%%%%%%%%%%%%%%%%%%%%%%%%%%%%%%%%%%%%%%%%%%%%%%%%%

\section{Pressure dependence of correlation}\label{sec:cor_press}
\begin{figure*}
\subfigure[]{\includegraphics[scale=0.33,angle=-90]{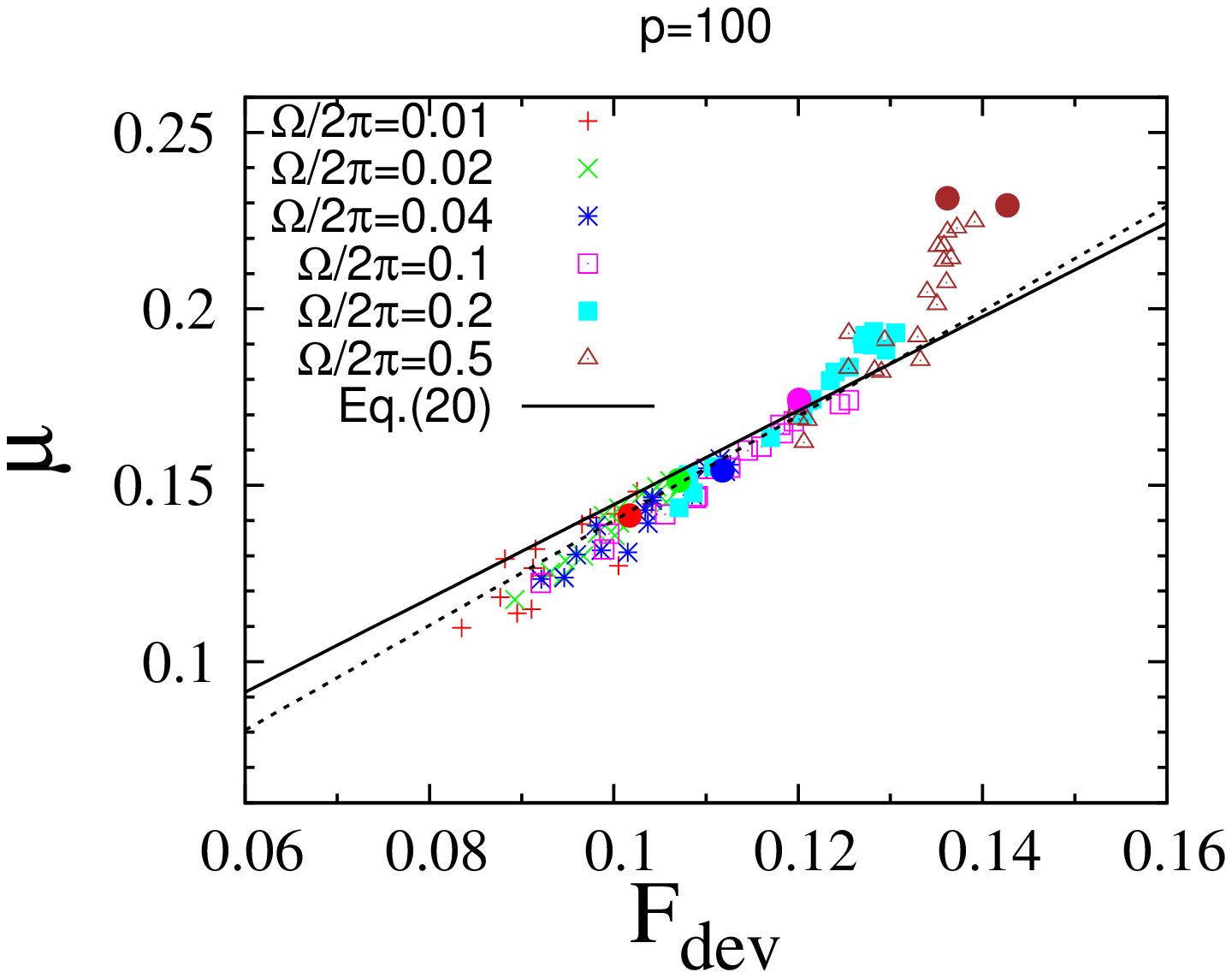}\label{fig:cor_press-100}}
\subfigure[]{\includegraphics[scale=0.33,angle=-90]{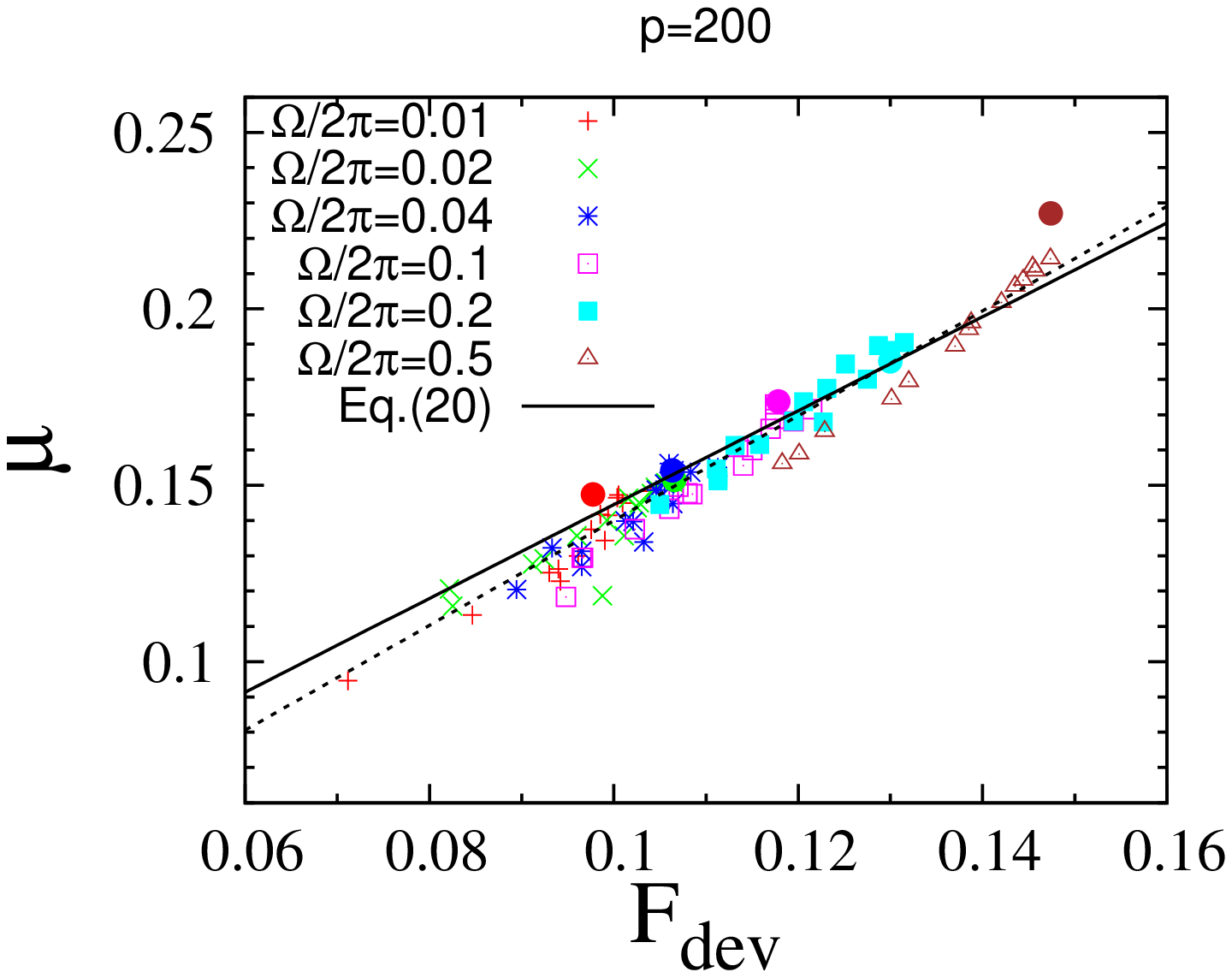}\label{fig:cor_press-200}}
\subfigure[]{\includegraphics[scale=0.33,angle=-90]{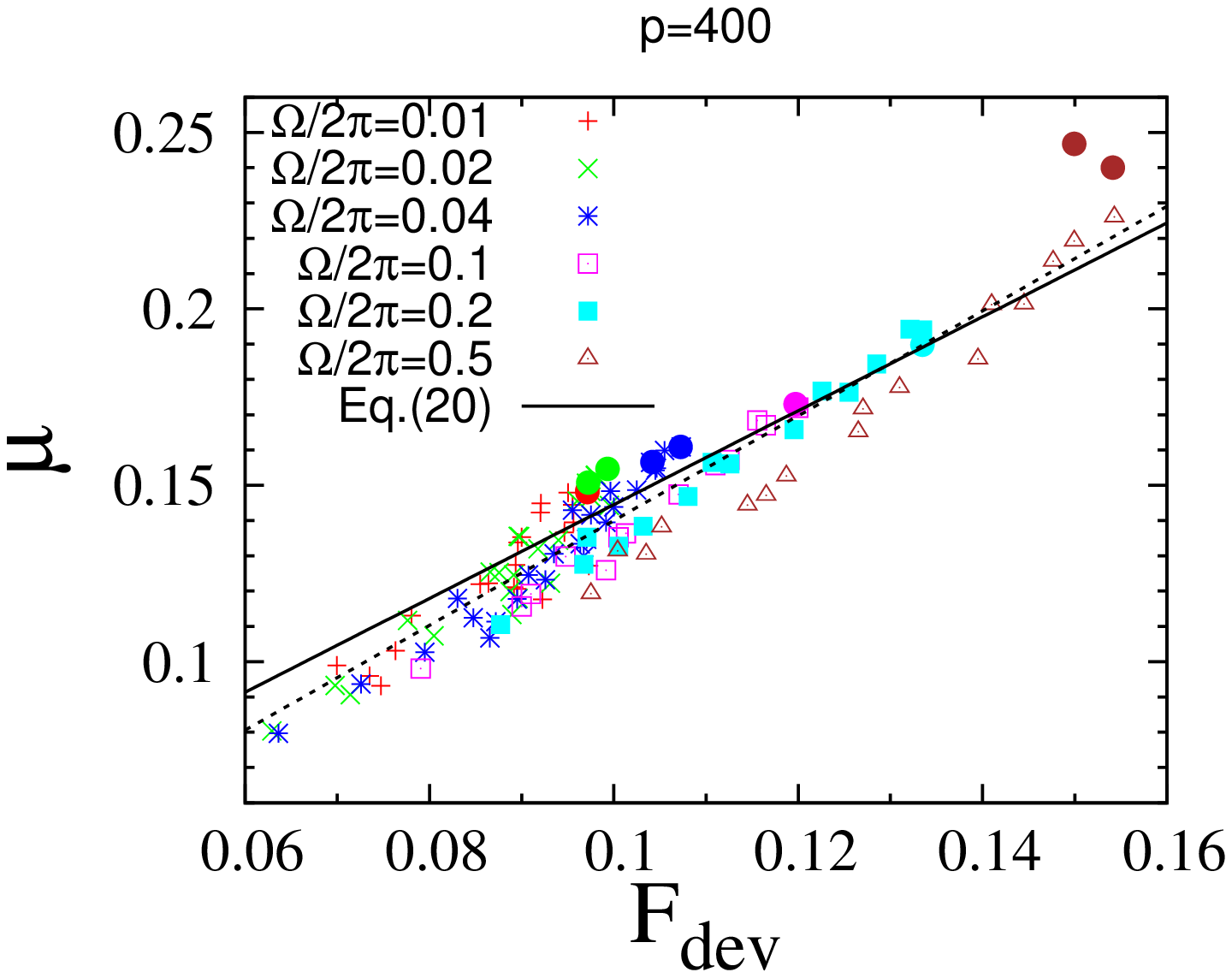}\label{fig:cor_press-400}}
\caption{(Color online)
$\mu$ plotted against $F_{\rm dev}$ for different local pressures in the system (a) $p=100$, (b) $p=200$, and (c) $p=400$ $\mathrm{Nm^{-2}}$.
The solid line represent the corresponding fit to \eref{eq:mu_local-F_dev}, while the dashed line is the best fit to the data.}
\label{fig:cor_press}
\end{figure*}

In this section, we test the correlation between $\mu$ and $F_{\rm dev}$ that was presented in section \ref{sec:iner}. 
Figure \ref{fig:cor_press} shows the data for three different pressure levels, we observe that the correlation holds
 very well for all rotation rates, except for the fastest rotation rate, which seems to fall off from the prediction of \eref{eq:mu_local-F_dev}.
%%%%%%%%%%%%%%%%%%%%%%%%%%%%%%%%%%%%%%%%%%%%%%%%%%%%%%%%%%%%%%%%%%%%%%%%%%%%%%%%

\section*{References}

\bibliographystyle{unsrt}

\bibliography{DiffgLib}

\end{document}